\documentclass[preprint,12pt]{elsarticle}

\usepackage{multirow}
\usepackage{amssymb}
\usepackage{mathtools}
\usepackage[ruled,vlined]{algorithm2e}
\usepackage{stmaryrd}
\usepackage{tikz}
\definecolor{color0}{rgb}{0.215686,0.494118,0.721569}
\definecolor{color1}{rgb}{1,0.498039215686275,0.0549019607843137}
\definecolor{color2}{rgb}{0.301961,0.686275,0.290196}
\definecolor{color3}{rgb}{0.83921568627451,0.152941176470588,0.156862745098039}
\definecolor{pltgreen}{rgb}{0,0.5,0}

\usetikzlibrary{decorations.pathmorphing}
\usetikzlibrary{snakes}
\usetikzlibrary{shapes}

\usepackage{pdflscape}
\usepackage{afterpage}
\usepackage{capt-of}
\usepackage{booktabs}

\usepackage{pgfplots}
\usepgfplotslibrary{groupplots,dateplot}
\usetikzlibrary{patterns,shapes.arrows}
\pgfplotsset{compat=newest}

\tikzset{cross/.style={cross out, draw=black, minimum size=2*(#1-\pgflinewidth), inner sep=0pt, outer sep=0pt},
cross/.default={2pt}}

\usepackage{siunitx}

\newcommand\pder[2][]{\ensuremath{\frac{\partial#1}{\partial#2}}}
\newcommand\pdder[2][]{\ensuremath{\frac{\partial^2 #1}{\partial#2^2}}}

\usepackage{hyperref}

\journal{Journal of Computational Physics}

\begin{document}

\begin{frontmatter}



\title{ModalPINN : an extension of Physics-Informed Neural Networks with enforced truncated Fourier decomposition for periodic flow reconstruction using a limited number of imperfect sensors}


\author[label1]{Gaétan Raynaud\corref{cor1}}
\ead{gaetan.raynaud@polymtl.ca}
\author[label2]{Sébastien Houde\corref{cor2}}
\ead{sebastien.houde@gmc.ulaval.ca}
\author[label1]{Frédérick P Gosselin\corref{cor3}}
\ead{frederick.gosselin@polymtl.ca}

\address[label1]{Département de Génie Mécanique, Laboratory for Multiscale Mechanics (LM2), Polytechnique Montréal, Montreal, QC, Canada}
\address[label2]{Hydraulic Machinery Laboratory LAMH, Faculty of Science and Engineering, Laval University, Québec, G1V 0A6, Canada}

\cortext[cor1]{Corresponding authors at: Department of Mechanical Engineering, Ecole Polytechnique, Montreal, QC, Canada.}

\begin{abstract}

Continuous reconstructions of periodic phenomena provide powerful tools to understand, predict and model natural situations and engineering problems. In line with the recent method called Physics-Informed Neural Networks (PINN) where a multi layer perceptron directly approximates any physical quantity as a symbolic function of time and space coordinates, we present an extension, namely ModalPINN, that encodes the approximation of a limited number of Fourier mode shapes. In addition to the added interpretability, this representation performs up to two orders of magnitude more precisely for a similar number of degrees of freedom and training time in some cases as illustrated through the test case of laminar shedding of vortices over a cylinder. This added simplicity proves to be robust in regards to flow reconstruction using only a limited number of sensors with asymmetric data that simulates an experimental configuration, even when a Gaussian noise or a random delay is added, imitating imperfect and sparse information.

\end{abstract}

\begin{graphicalabstract}
\include{images/tikz_fig_GraphicalAbstract}
\end{graphicalabstract}

\begin{highlights}
\item We propose a new architecture for Physics-Informed Neural Networks (PINN) specialised for oscillatory phenomena.
\item The hard-coded truncated Fourier decomposition allows a better precision than the classical approach at an equivalent number of degrees of freedom and computing time.
\item The proposed format shows great convergence for field reconstruction with data from a limited number of sensors and can overcome problems of time synchronisation and noise.
\end{highlights}


\end{frontmatter}


\section{Introduction}
\label{sec:introduction}

Data-assimilation techniques helped bridge the gap between experimentation, numerical simulation, and modelling in order to design better engineering solutions. However when data is expensive to gather and therefore scarce, these conventional solutions lack physical accuracy. New algorithms such as Physics-Informed Neural Networks that are presented in this paper allow using prior knowledge from Partial Differential Equations (PDE) to reconstruct continuous fields and predict quantities of interest in the small-data regime. This leads the way to improve the design of mechanical systems, monitor operations and perform predictive maintenance to reduce operating cost and increase overall efficiency.\\

Along with a series of advances in various fields \cite{goodfellow_deep_2016} such as computer vision \cite{krizhevsky_imagenet_2017} and natural language processing \cite{bengio_neural_2003}, artificial intelligence has found numerous applications in fluid dynamics. Especially, deep learning took advantage of the massive amount of experimental data and high-fidelity simulations \cite{brunton_machine_2020} with applications in flow estimation \cite{guo_convolutional_2016}, active control \cite{rabault_artificial_2019} or complex optimisation like collective swimming \cite{verma_efficient_2018}.\\

From the field of dynamical systems and sparse regression \cite{brunton_discovering_2016,corbetta_application_2020}, another branch has emerged and aims at generalising the idea of test functions using multi layer perceptrons that directly approximate any physical quantity as a symbolic function of spatial and temporal coordinates. This technique, called Physics-Informed Neural Networks (PINN) finds its origins in the early work of Dissanayake et al. \cite{dissanayake_neural-network-based_1994} and Lagaris et al. \cite{lagaris_artificial_1998}. It was met with renewed interest since 2018. PINNs allow the reconstruction of hidden variables using different types of data without preliminary processing, identification of PDE parameters and resolution of complex direct and inverse problems governed by these types of equations \cite{raissi_physics-informed_2019}.\\


The conciseness of implementation of this method and the profusion of PDE-governed phenomena has lead to numerous works applying the concepts of PINNs to diverse fields such as fluid dynamics with non-newtonian fluids \cite{reyes_learning_2020} and high-speed flows \cite{mao_physics-informed_2020}, but also in material sciences \cite{haghighat_deep_2020,shukla_physics-informed_2020}, electromagnetism \cite{hu_deep_nodate} or nano-optic and meta-materials \cite{chen_physics-informed_2020}.\\

Improvements to the mathematical basis of PINNs have been proposed to increase the efficiency and robustness of training. For instance our paper uses a technique called prior-dictionary \cite{peng_accelerating_2020} that allows enforcing prior knowledge about the solution such as boundary conditions. Other papers discuss techniques to improve the activation functions by adding degrees of freedom \cite{jagtap_adaptive_2019}, or address deeper issues like an adaptive weighting of loss parameters \cite{wang_when_2020} and gradient-related issues in the optimisation process \cite{wang_understanding_2020}.\\

Modal approaches have been a matter of interest in flow modelling. They allow lighter representations by extracting physically important patterns from raw data obtained by numerical simulation or experimentation \cite{taira_modal_2017}. This thematic has been addressed under the scope of machine learning with, for instance, Fourier content that is learned from the geometry in order to improve prediction performances during the design phase \cite{zhang_frequency-compensated_2020}. Spectral methods for high randomness have been enforced in PINNs governed by stochastics PDE \cite{zhang_learning_2019}. \\

Our paper is positioned in the continuity of Raissi et al. \cite{raissi_deep_2019} where a vortex-induced vibration (VIV) phenomena is modelled using a classical PINN. Inspired by harmonic balance techniques (HBT), we aimed at directly enforcing this oscillatory phenomena in the way PINN represents information so that it gains in interpretability. For simplicity reasons, structural movement has not been considered in the presented results and only the fluid flow has been reconstructed.\\

From an experimental point of view, flow reconstruction might be a difficult and expensive challenge. Some techniques make it possible to obtain flow information at discrete points in a volume or on a plane with Particle Image Velocimetry (PIV) and its tomographic and holographic variants. Other techniques give only information on traverses such as Laser Doppler Velocimetry (LDV) or at single points like pitot probe or hot-wire anemometry \cite{bailly_experimental_2015}. These techniques require a substantial time for calibration and may not be available everywhere or at the same time. Moreover, recording with one pitot probe at several location results in a set of asynchronous data. In the area close to a wall, the large gradients of velocity and the heat-loss through the wall create flaws in PIV \cite{kahler_uncertainty_2012} and hot-wire measurements \cite{durst_experimental_2002}. Some complex geometries like inter-blade regions in hydraulic turbines can be difficult to access and limit the area of visualisation with optical techniques \cite{aeschlimann_inter-blade_2013}. Other defects can also appear with the tracking particles such as peak-locking in PIV \cite{cholemari_modeling_2007} which can corrupt the data. In this context, PINNs can help fill the gap by interpolating between sparse data like plane measurements to reconstruct 3D fields from 2D measurements \cite{cai_flow_2021}. There are still some questions about the ability of PINN to deal with and correct these imperfections and to extrapolate outside of the available measurement window.\\

The objective of this paper is to propose a simpler representation of PINNs for oscillating phenomena, namely a ModalPINN, and quantitatively show that this simplification provides robustness regarding sparsity, noise and lack of synchronisation in the provided data. The following section recalls the mathematical grounds of PINNs and present our ModalPINN. The vortex-shedding that serves as a test case is introduced in the third section with the required adaptations. Then several training configurations are run from dense and perfect data to flow reconstruction using sparse and artificially corrupted time signals.

\section{Method}
\label{sec:method}

\subsection{Theoretical background about physics-informed neural networks}

ModalPINN is based on the concept of physics-informed neural network (PINN). Its formulation is presented in the next subsection alongside some precision on the use of prior-dictionary to enforce boundary conditions and unsteady force computation with PINN.

\subsubsection{Physics-Informed Neural Networks}

We consider a physical problem where an unknown variable $\mathbf{q}(\mathbf{x},t) \in \mathbb{R}^n$ is defined as a solution of a partial differential equation. This variable $\mathbf{q}$ is a function defined on a spatial domain $\Omega$ and on a time interval $[t_0,t_f]$. The set of equations also contains a boundary term on $\partial \Omega$ and initial conditions:
\begin{equation}
    \left\{
    \begin{array}{rll}
        \mathcal{N}(\mathbf{q},t) = & \mathbf{f}(\mathbf{x},t) & \forall \mathbf{x},t \in \Omega \times [t_0,t_f], \\
        \mathbf{q}(\mathbf{x},t) = & \mathbf{h}(\mathbf{x},t) & \forall  \mathbf{x},t \in \partial \Omega \times [t_0,t_f], \\
        \mathbf{q}(\mathbf{x},t_0) =& \mathbf{q}_0(\mathbf{x}) & \forall \mathbf{x} \in \Omega,
    \end{array}
    \right.
\end{equation}

\noindent where $\mathcal{N}$ is a differential operator with respect to spatio-temporal coordinates, and which can be non linear. \\

The idea behind a PINN is to approximate the physical solution $\mathbf{q}$ with a neural network. The neural network (NN) defined by its set of parameters $\theta \in \mathbb{R}^p$ is considered as a function of physical coordinates (here space and/or time). The approximated solution $\tilde{\mathbf{q}}$ is obtained with

\begin{equation}
    \tilde{\mathbf{q}} (\cdot) = NN(\theta;\cdot) \approx \mathbf{q}(\cdot),
\end{equation}

\noindent and is completely specified once all parameters $\theta$ are set. In other words, the approximation is continuously defined without any mesh required. For the purpose of concision, the tilde is dropped and $\tilde{\mathbf{q}}$ is referred as $\mathbf{q}$ from here on. It can also be noted that having time as one of the input coordinates $(\mathbf{x},t)$ is strictly equivalent as having an additional spatial dimension in $\mathbf{x}$.\\

The neural network $NN(\theta;\cdot)$ designates a symbolic graph of operations consisting of, alternatively, a matrix-vector product and a sum, and a non-linear activation function $\sigma : \mathbb{R}^j \rightarrow \mathbb{R}^j$. For a neural networks of depth $k$ defined with the set of parameters $\theta = \{W_0,\mathbf{b_0}, ... W_k, \mathbf{b_k}\}$, where $W_i$ are matrices and $\mathbf{b_i}$ vectors, one can obtain from an input $(\mathbf{x},t)$ the output $\mathbf{q}$ with the following sequence of operations

\begin{equation}
    \begin{array}{rcl}
        \mathbf{y_0} & = & \left( \mathbf{x},t \right) \in \mathbb{R}^{n_0}\text{, usually } n_0 =3 \text{ or } 4, \\
        \mathbf{y}_1 & = & \sigma\left( W_0 \mathbf{y_0} + \mathbf{b_0} \right) \in \mathbb{R}^{n_1}, \\
        & \vdots & \\
        \mathbf{y}_{i+1} & = & \sigma\left( W_i \mathbf{y_i} + \mathbf{b_i} \right)  \in \mathbb{R}^{n_{i+1}},\\
        & \vdots & \\
        \mathbf{q} & = & W_k \mathbf{y_k} + \mathbf{b_k}  \in \mathbb{R}^{n_{k+1}}.
    \end{array}
\end{equation}

\begin{figure}
    \centering
    \begin{tikzpicture}[thick,scale=0.7, every node/.style={scale=0.7}]

\tikzstyle{unit}=[draw,shape=circle,minimum size=1.1cm] 
\small
\node[unit](x) at (-0.5,5){$x$};
\node[unit](y) at (-0.5,3){$y$};
\node[unit](t) at (-0.5,1){$t$};

\node[unit](s11) at (2,6){$y_1^1$};
\node[unit](s21) at (2,4){$y_1^2$};
\node(dots) at (2,2){\vdots};
\node[unit](s31) at (2,0){$y_1^{n_1}$};

\draw[->](x) -- (s11);
\draw[->](x) -- (s21);
\draw[->](x) -- (s31);
\draw[->](y) -- (s11);
\draw[->](y) -- (s21);
\draw[->](y) -- (s31);
\draw[->](t) -- (s11);
\draw[->](t) -- (s21);
\draw[->](t) -- (s31);

\node[unit](s12) at (5,6){$y_2^1$};
\node[unit](s22) at (5,4){$y_2^2$};
\node(dots) at (5,2){\vdots};
\node[unit](s32) at (5,0){$y_2^{n_2}$};

\draw[->](s11) -- (s12);
\draw[->](s11) -- (s22);
\draw[->](s11) -- (s32);
\draw[->](s21) -- (s12);
\draw[->](s21) -- (s22);
\draw[->](s21) -- (s32);
\draw[->](s31) -- (s12);
\draw[->](s31) -- (s22);
\draw[->](s31) -- (s32);

\node(dotsa) at (6.5,6){$\ldots$};
\node(dotsb) at (6.5,4){$\ldots$};
\node(dotsc) at (6.5,0){$\ldots$};

\node[unit](s13) at (8,6){$y_k^1$};
\node[unit](s23) at (8,4){$y_k^2$};
\node(dots) at (8,2){\vdots};
\node[unit](s33) at (8,0){$y_k^{n_k}$};

\node[unit](v) at (10,3){$q$};

\draw[->](s13) -- (v);
\draw[->](s23) -- (v);
\draw[->](s33) -- (v);

\draw[decorate,decoration={brace,raise=-0.3,amplitude=10pt}] (1.2,7) -- (8.8,7);
\node at (5,8) {DNN hidden layers with parameters $\theta$};

\draw[decorate,decoration={brace,amplitude=10pt}] (-1.5,0.5) -- (-1.5,5.5);
\node[align=right] at (-3.8,3) {Physical \\ coordinates \\ $(x,y,t) \in \Omega \times [t_0,t_f] $};

\node[align=left] at (11.5,3.) {Physical \\ field $ \in \mathbb{R}$};

\end{tikzpicture}
    \caption{Physics Informed Neural Network classical structure for approximating a field $q$ as a function of spatio-temporal coordinates $(x,y,t)$. If $q$ were to be a vector, for instance velocity components and pressure $(u,v,p)$, all these quantities could go along in the output of one PINN or in separated neural networks.}
    \label{fig:PINN_structure}
\end{figure}
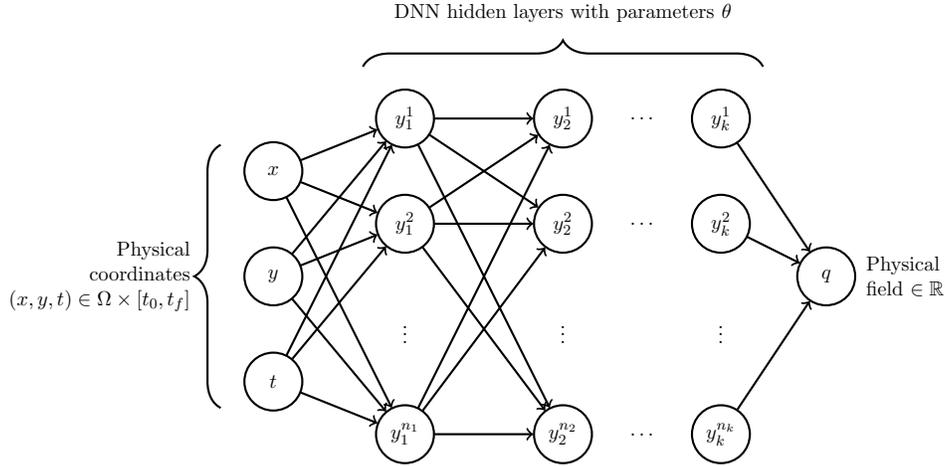

This sequence of operation is usually illustrated by a graph, as depicted in Figure \ref{fig:PINN_structure}. In this example, unknown quantities of a two dimensional incompressible flow $\mathbf{q} = (u,v,p)$ defined on a 2D cartesian domain $\mathbf{x} = (x,y)$, are solved using a PINN for each scalar variable. It is also possible to unite all the flow quantities in the same PINN. The choice of activation function and neural network size are both yet to be settled in the literature dealing with PINN. A comparison can be made with numerical methods such as finite elements where the number of degrees of freedom is linked to the number of parameters $\theta$ that quantifies the network's size. Besides, activation function $\sigma$ can be viewed as a form function that will help approximate any complicated shape. For classical cases, sine and hyperbolic tangent proved to work in previous studies \cite{raissi_deep_2019,raissi_deep_2018}. We adopted the same choice by using $\sigma=\sin$ when there is periodicity with one of the input coordinates, and $\tanh$ in other cases like mode shapes reconstruction. \\

Typical PINN algorithms optimise the set of parameters $\theta$ in order to minimise a specific loss function $\mathcal{L}$. For a PINN, the loss function is generally composed of two kinds of terms: $\mathcal{L} = \mathcal{L}_m + \mathcal{L}_{eq}$ where $\mathcal{L}_m$ and $\mathcal{L}_{eq}$ respectively represent:
\begin{enumerate}
    \item The distance to measurements or Dirichlet boundary conditions. On a sample of coordinates $V_m$ of size $N_m$, $\mathcal{L}_m$ represents the average squared distance to specific and known values $\mathbf{q}_m$:
    \begin{equation}
        \mathcal{L}_m = \frac{1}{N_m} \sum_{\mathbf{x}_m,t_m \in V_m} \left| \tilde{\mathbf{q}}(\mathbf{x_m},t_m) - \mathbf{q}_m \right|^2,
    \end{equation}
    Using a quadratic norm allows smoother differentiation. Here $\mathbf{q}_m$ can be a sampling of measurements data as well as boundary conditions. In that last case, $V_m$ would be a discrete sampling of $\partial \Omega \times I$ with $\mathbf{q}_m = h(\mathbf{x}_m,t_m)$.
    \item The residuals of partial differential equations or Neumann boundary conditions:
    \begin{equation}
    \label{eq:Loss_eqs_global}
        \mathcal{L}_{eq} = \frac{1}{N_{in}} \sum_{\mathbf{x}_{in},t_{in} \in V_{in}} \left| \mathcal{N}(\tilde{\mathbf{q}}(\mathbf{x}_{in},t_{in})) - f(\mathbf{x}_{in},t_{in}) \right|^2,
    \end{equation}
    where $V_{in}$ is a sampling of the PDE domain $\Omega \times [t_0,t_f]$ where $\mathbf{q}$ is defined, and $N_{in}$ its cardinal. For Neumann boundary conditions, $V_{in}$ would be a sampling of $\partial \Omega \times [t_0,t_f]$ and $\mathcal{N}$ and $f$ would be adapted consequently.
\end{enumerate}
\medskip

The second part benefits from automatic differentiation available with neural networks. Since every operation in the operation graph is known and differentiable, derivatives with respect to any variable in the graph can be computed exactly with most machine learning libraries such as TensorFlow \cite{abadi_tensorflow_2016}. Most of the time, machine learning makes use of this property to perform fast optimisation of parameters $\theta$ (with gradient descent for instance). But since the input has a physical signification in PINN, it makes sense to differentiate with respect to one of the input to compute gradients of the solution in physical space.\\

Once the loss function and sampling spaces $V_m$ and $V_{in}$ are defined, the model's parameters $\theta$ are optimised so that the approximated solution $\mathbf{q}$ fits best both equations and measurements. Several minimising algorithms are available. The quasi-Newtonian L-BFGS-B \cite{byrd_limited_1995} followed by the Adam optimisers \cite{kingma_adam_2017} are used and seemed effective from empirical observations. Technical details about training can be found in section \ref{sec:TechnicalDetails}.\\

At the end of the training, $\mathcal{L}_m$ is computed using a larger data set $V_m^{\text{valid.}}$ from simulations with new points that have not been used for optimisation. This provides a squared $L^2$ measure of the reconstruction error that is later referred as validation error. It should be noted that outside of PINNs literature, this quantity may be named testing loss in the field of machine learning.\\

\subsubsection{Prior-dictionary}

One way to take the boundary conditions into account is to penalise the error on a sampling of points. This is the method presented in the previous section and which is included in the $\mathcal{L}_m$ term. Nonetheless, it may slow down or prevent the algorithm from converging on a solution. This issue has been tackled by Peng et al. \cite{peng_accelerating_2020} who proposed a method called prior-dictionary. \\

The idea is to force the shape of the approximated solution to fit some criteria, especially Dirichlet conditions. If the condition to be satisfied is $\mathbf{q}(\mathbf{x},t) = h(\mathbf{x},t), \forall \mathbf{x} \in \partial \Omega$ which is independent of time, the approximated solution can be defined as
\begin{equation}
\label{eq:PriorDictionnary}
    \tilde{\mathbf{q}}(\mathbf{x},t) = NN(\theta;\mathbf{x},t)\times f_{BC} (\mathbf{x}) + h(\mathbf{x},t),
\end{equation}

\noindent using a function $f_{BC}$ which equals 0 at specific domain frontiers $\partial \Omega$. Following the example of a two-dimensional flow where $\mathbf{q} = (u,v)$ and $\mathbf{x} = (x,y)$, it is possible to impose a no-slip boundary condition on $y=0$ by defining $f_{BC}(x,y) = \tanh{y}$. This choice is not unique. The difficulty is to select a $f_{BC}$ that is almost flat on the entire domain except at specific boundaries in order to minimise the deformation of the neural network output $NN(\theta;\mathbf{x},t)$. Recent works \cite{lu_physics-informed_2021,rao_hard_2021,sukumar_exact_2021} have demonstrated the effectiveness of this approach. Therefore it is used for all the presented results.

\subsubsection{Unsteady force computations}

For the computation of forces on any border, a parametric definition is used. For instance a 1D-frontier in a two-dimensional domain can be defined by $s\in [0,1] \rightarrow \left(x_{BC}(s),y_{BC}(s)\right) \in \mathbb{R}^2$. This is a symbolic function, either analytically defined by equations (as lines, circles, parabola...) but it can also be defined by an auxiliary neural network, pre-trained to fit any regular border. This allows a PINN to use more complex border shapes. Besides, this parametric approach can be generalised to higher dimensions, such as a surface defined by $(s,\xi) \in [0,1]^2 \rightarrow (x_{BC},y_{BC},z_{BC}) \in \mathbb{R}^3$. Moreover for non canonical shapes, for instance with discontinuities, it is possible to define several borders that can be separately approximated by a symbolic function.\\

Once a symbolic function of the border is available, computation of the normal vector is made possible using the automatic differentiation of neural networks with respect to $s$. In a two-dimensional problem, the normal vector is given by:
\begin{equation}
    \vec{n}(s) = \left(n_x(s),n_y(s)\right) = \left( - \pder[y_{BC}]{s}(s), \pder[x_{BC}]{s}(s) \right).
\end{equation}

\noindent Then, the total forces $\vec{F}$ on a border can be estimated using an empirical average with the Monte Carlo method in order to integrate local forces $\vec{df}(x,y)$:

\begin{equation}
\label{eq:integral_force}
    \vec{F} = \int_{\partial \Omega_f} \vec{df} dl.
\end{equation}

\noindent In the case of a two-dimensional incompressible flow, local forces $\vec{df} = \left(df_x,df_y \right)$ are given by

\begin{align}
    df_x =& -pn_x + \frac{2}{Re}\pder[u]{x}n_x + \frac{1}{Re}\left( \pder[u]{y} + \pder[v]{x}\right)n_y, \\
    df_y =& -pn_y + \frac{2}{Re} \pder[v]{y} n_y + \frac{1}{Re} \left( \pder[u]{y} + \pder[v]{x} \right)n_x,
\end{align}

\noindent where $Re$ is the Reynolds number quantifying the ratio between inertial and viscous forces, $u,v$ and $p$ are the dimensionless velocity and pressure fields. Then the integral in equation (\ref{eq:integral_force}) is approached by the symbolic parametrization and a Monte-Carlo method

\begin{equation}
\begin{split}
\vec{F}(t) =& \int_{\partial \Omega_f} \vec{df}(x,y,t)dl = \int_{[0,1]} \vec{df}\left(x_{BC}(s),y_{BC}(s),t\right) \left| \frac{dl(s)}{ds} \right| ds, \\
\approx & \frac{1}{\text{card}V_s} \sum_{s \in V_s} \vec{df}\left(x_{BC}(s),y_{BC}(s),t\right) \left| \frac{dl(s)}{ds} \right|,
\end{split}
\end{equation}

\noindent where $V_s$ is a sampling of $[0,1]$ which is then mapped to the coordinates of the points on the boundary using the parametrization $s \rightarrow x_{BC}(s),y_{BC}(s)$. This sampling $V_s$ can be uniform, in which case $s$ is the curvilinear abscissa divided by the length of the border $L$ and $\left| \frac{dl(s)}{ds} \right| = L$. But in case of strong variations in the integrand, an adaptive sampling can be used with Monte-Carlo method. In that case, $\left| \frac{dl(s)}{ds} \right|$ is calculated using the probability distribution function of the random variable $s$.

\subsection{ModalPINN : enforcing Fourier modes in the neural architecture}

Periodicity occurs for a wide range of phenomena in nature and in engineering processes. The mathematical tools and models can be adapted to use this property to significantly speed-up calculations. The following subsections present how a truncated modal representation can be directly included into the neural network architecture and how it allows another type of physical regularisation based on modal equations.

\subsubsection{Modal decomposition encoded in ModalPINN}

For a physical case where the observed phenomena is periodic for one space-time coordinate, it can be convenient to decompose the solution with a modal approach. For example, consider a real function of space and time $q(\mathbf{x},t)$ periodic in time with a fundamental frequency $f_0 = 2\pi \omega_0$. It can be transformed with Fourier decomposition such as:
\begin{equation}
    q(\mathbf{x},t) = \sum_{k=0}^\infty \hat{q}_k(\mathbf{x})e^{ik\omega_0 t} + c.c.,
\end{equation}
\noindent where $c.c.$ designates the complex conjugate and $\hat{q}_k \in \mathbb{C}$ are the modal coefficients at frequency $2\pi k \omega_0$ with $k\in\mathbb{N}$. These coefficients are functions of space only, which removes time as a variable needed to solve the problem.\\

In some circumstances, it is possible to obtain an acceptable approximation of $q(\mathbf{x},t)$ with a finite number of modes. The obtained level of accuracy may depend on the presence of high frequency phenomena. Besides, high order harmonics may be required when non-linear features in the governing equations lead to interactions between modes at different frequencies. Given a number of modes $N$, a PINN with prior dictionaries aiming at approximating these modal shapes is constructed:
\begin{equation}
    \mathbf{x} \in \Omega \xrightarrow[]{NN(\theta;\cdot) \times f_{BC}(\cdot)} (\hat{q}_0, ..., \hat{q}_N) \in \mathbb{C}^{N+1}.
\end{equation}

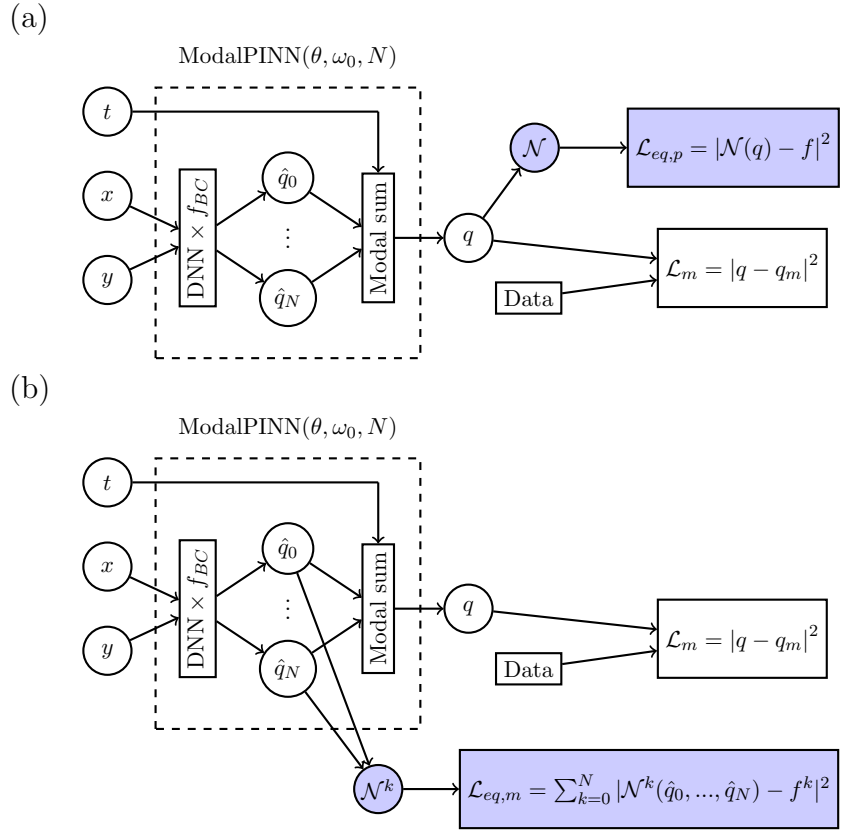
\begin{figure}
    \centering

    \begin{tabular}{ll}
        (a) & \\
        & \begin{tikzpicture}[thick,scale=0.8, every node/.style={scale=0.8},cross/.style={path picture={ 
  \draw[black]
(path picture bounding box.south east) -- (path picture bounding box.north west) (path picture bounding box.south west) -- (path picture bounding box.north east);
}}]

\tikzstyle{unit}=[draw,shape=circle,minimum size=0.8cm] 
\small
\node[unit](x) at (0,3.7){$x$};
\node[unit](y) at (0,2.3){$y$};

\node[rectangle,draw,rotate=90](DNN) at (1.5,3){$\text{DNN}\times f_{BC}$};
\draw[->](x) -- (DNN);
\draw[->](y) -- (DNN);

\node[unit](u0) at (3,4){$\hat{q}_0$};
\node(dots) at (3,3.1){\vdots};
\node[unit](un) at (3,2){$\hat{q}_N$};

\draw[->](DNN) -- (u0);
\draw[->](DNN) -- (un);

\node[unit](t) at (0,5.1){$t$};

\node[rectangle,draw, rotate=90](ModSum) at (4.5,3){Modal sum};

\draw[->](u0) -- (ModSum);
\draw[->](un) -- (ModSum);

\draw (t) -- (4.5,5.1);

\draw[->] (4.5,5.1) -- (ModSum);

\node[unit](output) at (6,3){$q$};
\draw[->](ModSum) -- (output);

\draw[draw=black,dashed](0.8,1) rectangle (5.2,5.5);
\node[align=center] at (3,6){ModalPINN$(\theta,\omega_0,N)$};

\node[unit,inner sep=1pt,,fill=blue!20](N) at (7.1,4.5){$\mathcal{N}$};

\draw[->](output) -- (N);

\node[rectangle,draw,align=center,minimum height = 1.3cm,fill=blue!20](LossEqsGlobales) at (10.4,4.5){$\mathcal{L}_{eq,p} = \left| \mathcal{N}(q) - f \right|^2$};
\draw[->](N) -- (LossEqsGlobales);

\node[rectangle,draw](Data) at (7,2){Data};
\node[rectangle,draw,align=center,minimum height = 1.3cm](LossMes) at (10.55,2.5){$\mathcal{L}_{m} = \left|q - q_{m} \right|^2$};
\draw[->](output) -- (LossMes);
\draw[->](Data) -- (LossMes);

\end{tikzpicture} \\
        (b) & \\
        & \begin{tikzpicture}[thick,scale=0.8, every node/.style={scale=0.8},cross/.style={path picture={ 
  \draw[black]
(path picture bounding box.south east) -- (path picture bounding box.north west) (path picture bounding box.south west) -- (path picture bounding box.north east);
}}]

\tikzstyle{unit}=[draw,shape=circle,minimum size=0.8cm] 
\small
\node[unit](x) at (0,3.7){$x$};
\node[unit](y) at (0,2.3){$y$};

\node[rectangle,draw,rotate=90](DNN) at (1.5,3){$\text{DNN}\times f_{BC}$};
\draw[->](x) -- (DNN);
\draw[->](y) -- (DNN);

\node[unit](u0) at (3,4){$\hat{q}_0$};
\node(dots) at (3,3.1){\vdots};
\node[unit](un) at (3,2){$\hat{q}_N$};

\draw[->](DNN) -- (u0);
\draw[->](DNN) -- (un);

\node[unit](t) at (0,5.1){$t$};

\node[rectangle,draw, rotate=90](ModSum) at (4.5,3){Modal sum};

\draw[->](u0) -- (ModSum);
\draw[->](un) -- (ModSum);

\draw (t) -- (4.5,5.1);

\draw[->] (4.5,5.1) -- (ModSum);

\node[unit](output) at (6,3){$q$};
\draw[->](ModSum) -- (output);

\draw[draw=black,dashed](0.8,1) rectangle (5.2,5.5);
\node[align=center] at (3,6){ModalPINN$(\theta,\omega_0,N)$};

\node[rectangle,draw](Data) at (7,2){Data};
\node[rectangle,draw,align=center,minimum height = 1.3cm](LossMes) at (10.55,2.5){$\mathcal{L}_{m} = \left|q - q_{m} \right|^2$};
\draw[->](output) -- (LossMes);
\draw[->](Data) -- (LossMes);

\node[unit,inner sep = 1pt,fill=blue!20](mNk) at (4.5,0.){$\mathcal{N}^k$};

\node[rectangle,draw,align=left,minimum height = 1.3cm,fill=blue!20](LossEqsModales) at (9.,0){$\mathcal{L}_{eq,m}= \sum_{k=0}^N |\mathcal{N}^k(\hat{q}_0,...,\hat{q}_N)-f^k|^2$};
\draw[->](mNk) -- (LossEqsModales);
\draw[->](u0) -- (mNk);
\draw[->](un) -- (mNk);
\end{tikzpicture}
    \end{tabular}

    \caption{Schematic structure of the ModalPINN with access to mode shapes as well as spatio-temporal solution to compute partial-differential equations: (a) in the physical space; (b) in modal space.}
    \label{fig:ModalPINN_Structure}
\end{figure}

\noindent The complete approximated solution is recovered by the sum :
\begin{equation}
    \tilde{q}(\mathbf{x},t) = 2 Re \left(\sum_{k=0}^N \hat{q}_k (\mathbf{x}) e^{ik\omega_0 t}\right),
\end{equation}
\noindent all this can be done in the computational graph of the neural network, as illustrated in Figure \ref{fig:ModalPINN_Structure} and summarised in Algorithm \ref{algo:ModalPINN}. For vector-valued functions, the decomposition is applied to each scalar-valued component alike thanks to linearity of the sum.\\

\begin{algorithm}
\caption{Algorithm to create and train a ModalPINN}
\SetAlgoLined
\KwIn{Hyper-parameters of the Neural Network (size, $\sigma$) and optimisation (method, learning rate $\eta$, training limit ...)}
\KwIn{Number of modes $N$ and fundamental frequency $\omega_0$}
\KwResult{Modal decomposition encoded in a PINN}
Construct the structure of a dense Neural Network $x,y;\theta \rightarrow NN(x,y;\theta)$\;
Apply Prior-Dictionary to compute mode shapes : $\hat{q}_0,\hat{q}_1,...,\hat{q}_N$\;
Construct the modal sum : $q(x,y,t) = \sum_{k=0}^N \hat{q}_k(x,y) e^{ik\omega_0 t}$\;
Construct the fitting part of the loss function $ \mathcal{L}_m\left[V_m;\theta\right]$\;
Construct equation penalisation loss $ \mathcal{L}_{eq}\left[V_{in};\theta\right]$\;
Construct the total loss function for training $\mathcal{L} = \mathcal{L}_m + \mathcal{L}_{eq}$\;
Prepare training data set $V_{in},V_{m}$\;
Initialise the parameters of the model $\theta$\;
\While{training limit is not reached}{
Prepare the batch $\tilde{V}_{in},\tilde{V}_m \subset V_{in},V_m $\;
Compute the loss $\mathcal{L}\left[\tilde{V}_{in},\tilde{V}_m;\theta\right]$\;
Compute loss derivatives  $\pder[\mathcal{L}]{\theta}\left[\tilde{V}_{in},\tilde{V}_m;\theta\right]$\;
Update parameters $\theta$ using the optimiser strategy\;
}
Compute loss on validation data\;
\label{algo:ModalPINN}
\end{algorithm}

\subsubsection{Loss construction with physical and modal equations}
\label{subseq:LossConstruction}

Since a modal sum can be considered as an auxiliary neural network, derivatives of $q$ with respect to time and space are available. Consequently, one direct way of computing a loss function to penalise equations residuals is to use the same formalism as in the classical PINN approach. Thus, the modal sum is used as input to the partial differential operator. In that case, sampling space $V_{in}$ provides a sampling in both space and time. For the penalisation to be satisfactory, the number of points required is significantly higher than for a space-only problem. Though, for a time periodic solution, the time domain can be reduced to $[0,\frac{2\pi}{\omega_0}]$. This will be referred to as physical equations.\\

To go beyond this space and time sampling, an advantage can be drawn from the availability of modal shapes. By projecting the equation on a basis of oscillatory function, one can obtain modal operators :

\begin{equation}
\label{eq:ModalOperator}
    \mathcal{N}^k (\hat{q}_0,...,\hat{q}_N,\omega_0) = \int_0^{2\pi/\omega_0} \mathcal{N}\left( \sum_{j=0}^N \hat{q}_j(\mathbf{x})e^{ij\omega_0 t} + c.c. \right) \times e^{-ik\omega_0 t} dt,
\end{equation}

\noindent as well as modal forces obtained with a similar projection of the global forcing $\mathbf{f}(x,t)$ on the $k^{th}$ frequency, as obtained for $\mathcal{N}^k$ in equation (\ref{eq:ModalOperator})

\begin{equation}
\label{eq:ModalForce}
    f^k(\mathbf{x}) = \int_0^{2\pi/\omega_0} \mathbf{f}(x,t)e^{-ik\omega_0 t}dt.
\end{equation}

\noindent The possibility is therefore given to optimise the model on physical equations residuals $\mathcal{L}_{eq,physical}$ as formulated in equation (\ref{eq:Loss_eqs_global}) or by using residuals of modal equations $\mathcal{L}_{eq,modal}$, as illustrated in Figure \ref{fig:ModalPINN_Structure}. This part of the loss function may be formulated as follows

\begin{equation}
\label{eq:Loss_eqs_modal}
    \mathcal{L}_{eq,modal} = \sum_{k=0}^{N} \frac{1}{N_{in}} \sum_{\mathbf{x}_{in} \in V_{in}}\left| \mathcal{N}^k(\hat{q}_0(\mathbf{x}_{in}),...,\hat{q}_N(\mathbf{x}_{in})) - \mathbf{f}^k(\mathbf{x}_{in}) \right|^2,
\end{equation}

\noindent and for the purpose of conciseness it will be referred as $\mathcal{L}_{eq,m}$ (and $\mathcal{L}_{eq,physical}$ as $\mathcal{L}_{eq,p}$).\\

\section{Laminar vortex-shedding around a cylinder : a non-linear test case for ModalPINN}

We consider a two dimensional incompressible flow over a cylinder, where non-linear vortex shedding is known to occur when a critical Reynolds number is reached. In its dimensionless form, the diameter $d=1$, the horizontal inflow is the typical velocity scale $(u_\infty,v_\infty)=(1,0)$. In the present case, the Reynolds number is set at $Re = 100$. In this regime, periodic oscillations of velocity $(u,v)$ and pressure $p$ happen at a Strouhal number $St = \frac{fd}{u_\infty} \approx 0.17$ as documented by Fey et al. \cite{fey_new_1998} (the relative error on the Strouhal with the proposed fitting is estimated at \num{3e-4}). Numerical data are provided by Boudina et al. \cite{boudina_vortex-induced_2020} for a 2D simulation of the incompressible flow over a fixed cylinder and are available for download \cite{boudina_numerical_2021}. These are obtained using the finite element solver \textsc{Cadyf} \cite{etienne_perspective_2009} that performs hp-adaptive backward differential formula methods \cite{hay_hp-adaptive_2015} in order to keep the local truncation error under a given threshold.\\

\begin{figure}
    \centering
    \begin{tabular}{ll}
        (a) &  \\
        \resizebox{0.97\linewidth}{!}{
    \input{images/tikz_fig_Comput_domain}
    }
    \end{tabular}
    \begin{tabular}{ll}
        (b) & (c) \\
        \resizebox{0.45\linewidth}{!}{\input{Python_data_figure/FBC5/figure_tikz_plot_fbc5}} &     \includegraphics[width=0.4\linewidth]{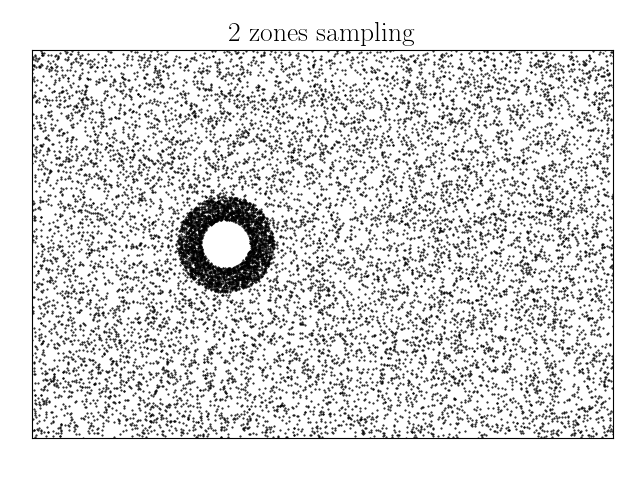}
    \end{tabular}

    \caption{(a) Computational domain and ModalPINN reduced domain (blue rectangle) used for training. (b) Prior-Dictionary $f_{BC}$ that enforces boundary conditions $u=v=0$ on the cylinder border. Profile of $f_{BC}$ along the centre line (dashed line) is plotted above. (c) Distribution of penalisation points for equations in 2 zones configuration with a number of points $N_{in}=15\times 10^3$.}
    \label{fig:Comput_Domain}
\end{figure}

Whereas the simulation domain in Boudina et al. \cite{boudina_vortex-induced_2020} is a rectangle of $(x,y) \in [-40,120]\times [-60,60]$, the domain used for the ModalPINN reconstruction covers a limited area defined by $(x,y) \in [-4,8]\times [-4,4]$ as depicted in Figure \ref{fig:Comput_Domain}a. In time, the simulation data used for reconstruction covers approximately 3 oscillation periods with 201 equally spaced time steps.\\

In order to impose boundary conditions on the cylinder, the following prior dictionary, as defined in equation (\ref{eq:PriorDictionnary}), is used for velocities $u$ and $v$
\begin{align}
    f_{BC}(x,y) &= \tanh \left[ \gamma \left( r-r_c \right)\right], \\
    h(x,y) &= 0,
\end{align}
\noindent where $r^2 = \left( x-x_c \right)^2 + \left(y-y_c\right)^2$ with $(x_c,y_c)=(0,0)$ being cylinder's coordinates and $r_c = 1/2$ its radius. The slope of $f_{BC}$ near the boundary is defined by the factor $\gamma$. In the present case $\gamma=5$ which is a compromise between a short transition zone and finite gradients. This function is depicted in Figure \ref{fig:Comput_Domain}b along its profile on centre line. \\

The equations which are to be solved by minimising the residuals are given by the three differential operators $\mathbf{\mathcal{N}} = \left(\mathcal{N}_{div},\mathcal{N}_x, \mathcal{N}_y \right)$ of the unknown $\mathbf{q} = (u,v,p)$. They stand for conservation of mass (\ref{eq:NS_global_div_u}) and momentum (\ref{eq:NS_global_f_u},\ref{eq:NS_global_f_v}):

\begin{align}
\label{eq:NS_global_div_u}
        \mathcal{N}_{div} (\mathbf{q}) = \pder[u]{x} + \pder[v]{y} & =  0, \\
\label{eq:NS_global_f_u}
        \mathcal{N}_x (\mathbf{q}) = \pder[u]{t} + u\pder[u]{x} + v\pder[u]{y} + \pder[p]{x} - Re^{-1}\left( \pdder[u]{x} + \pdder[u]{y} \right) & =  0,  \\
\label{eq:NS_global_f_v}
        \mathcal{N}_y (\mathbf{q}) = \pder[v]{t} + u\pder[v]{x} + v\pder[v]{y} + \pder[p]{y} - Re^{-1}\left( \pdder[v]{x} + \pdder[v]{y} \right) & =  0.
\end{align}

The associated modal operators for equations (\ref{eq:NS_global_div_u}), (\ref{eq:NS_global_f_u}) and (\ref{eq:NS_global_f_v}) are respectively noted as $\mathbf{\mathcal{N}}^k = \left(\mathcal{N}^k_{div},\mathcal{N}^k_x, \mathcal{N}^k_y \right)$. Modal representation of mass conservation writes as
\begin{equation}
\label{eq:ModalEquationsDivFree}
    \mathcal{N}^k_{div} = \pder[\hat{u}_{k}]{x} + \pder[\hat{v}_{k}]{y} , \forall k \in \llbracket 0 , N \rrbracket.
\end{equation}

Momentum balance along the $x$ axis of the $k^{th}$ mode writes as

\begin{equation}
\label{eq:ModalEquationsX}
    \begin{split}
            \mathcal{N}^k_x =& (ik\omega_0) \hat{u}_k + \pder[\hat{p}_k]{x} - Re^{-1}\left( \pdder[\hat{u}_k]{x} + \pdder[\hat{u}_k]{y} \right)
    + \sum_{l=0}^{k} \left( \hat{u}_l\pder[\hat{u}_{k-l}]{x} + \hat{v}_l \pder[\hat{u}_{k-l}]{y}\right) \\
    +& \sum_{l=k+1}^N \left( \hat{u}_l \pder[\hat{u}^*_{l-k}]{x} + \hat{u}^*_{l-k} \pder[\hat{u}_l]{x} + \hat{v}_l \pder[\hat{u}^*_{l-k}]{y} + \hat{v}^*_{l-k} \pder[\hat{u}_l]{y}\right), \forall k \in \llbracket 0 , N \rrbracket,
    \end{split}
\end{equation}

\noindent where $\hat{u}^*_k$ stands for the complex conjugate of the $k^{th}$ modal component of $u$. And similarly for the $y$ component of the momentum equation, the modal operator is obtained with

\begin{equation}
\label{eq:ModalEquationsY}
\begin{split}
    \mathcal{N}^k_y =&    (ik\omega_0) \hat{v}_k + \pder[\hat{p}_k]{y} - Re^{-1}\left( \pdder[\hat{v}_k]{x} + \pdder[\hat{v}_k]{y} \right)
    + \sum_{l=0}^{k} \left( \hat{u}_l\pder[\hat{v}_{k-l}]{x} + \hat{v}_l \pder[\hat{v}_{k-l}]{y}\right)  \\
    +& \sum_{l=k+1}^N \left( \hat{u}_l \pder[\hat{v}^*_{l-k}]{x} + \hat{u}^*_{l-k} \pder[\hat{v}_l]{x} + \hat{v}_l \pder[\hat{v}^*_{l-k}]{y} + \hat{v}^*_{l-k} \pder[\hat{v}_l]{y}\right), \forall k \in \llbracket 0 , N \rrbracket.
\end{split}
\end{equation}

Penalisation of equations is conducted on a randomly generated sampling of points $V_{in}$. Different strategies of space sampling may be defined. From the basic uniform sampling to a sampling adapted to the solution's local complexity, the final choice depends on a compromise between calculation speed and precision. A 2 zones sampling is used here. It consists in distributing 80 \% of points uniformly and concentrating the last 20 \% within a given distance around the cylinder, as depicted in Figure \ref{fig:Comput_Domain}c. By doing this, the relative weight of the residuals located in the boundary layer of the cylinder increases in comparison of those in the rest of the fluid domain. Thus, the shear layer, its detachment and near pressure field are expected to be more accurate which should lead to an increased precision in the estimation of the forces. This spatial sampling has been used for results presented in Figures \ref{fig:resultsFromSimulatedMeas_NMSE_Diff}, \ref{fig:resultsFromSimulatedMeas_Residuals_Forces}, \ref{fig:Distribution_residuals} and \ref{fig:Data_resyncro}. However, results depicted in Figures \ref{fig:ModalPINN_vs_PINN}, \ref{fig:ModeShapesEqsGlobalVsEqsModal} and \ref{fig:NoiseDependency} used a uniform spatial sampling optimised for the validation of the equations in the domain. Those two different strategies are also linked with limitations in the number of penalisation points due to memory overflow problems. But from a theoretical point of view, the results are expected to converge through a similar solution as the number of penalisation points increases thanks to larger and better distributed computational resources.\\

\section{Results}
\label{sec:results}

In this section, results on several training configurations are presented from the one with the largest training data to cases with sparse and flawed information. The first part aims at testing how a ModalPINN performs in comparison to a classical PINN. It also provides some insights about the use of modal equations. The last sections highlight the ability of ModalPINN to address ill-posed problems in simulated experimental conditions. For each result, run properties are recalled in Table 1.\\

\subsection{Comparison between ModalPINN and classical PINN approach}

To illustrate the simplicity brought by the ModalPINN, a comparison is performed with the classical PINN approach that approximates the entire field $x,y,t \rightarrow q(x,y,t)$ as a symbolic function of three coordinates. As the oscillatory nature of the phenomena is known, a  sine function is chosen as the activation function between layers to ease convergence, as it has been done by Raissi et al. \cite{raissi_deep_2019}. A training is performed with a time limit of 2 hours using equivalent computational resources (see section \ref{sec:TechnicalDetails} and Table 1). Physical equations are used for both the classical PINN and ModalPINN as well as hard encoding of boundary conditions with prior-dictionary. Time sampling for equations penalisation is performed over the simulation data range since the classic PINN is not able to extrapolate the periodic phenomena outside its trained time range.\\

In classical numerical simulations like finite elements, dependency of the precision of the solution with the size of the mesh is a key parameter to compare two algorithms. In a PINN, the similar quantity is the number of parameters to optimise. Their influence on precision have been examined for test purpose. To do so, the width of each hidden-layer is multiplied by a factor $W_l$ according to the structure detailed in Table 1. Training is performed using physical equations on a set of randomly sampled points in the domain (with uniform probability) and $N_m = 5000$ measurements $(u_m,v_m,p_m)$ at points $(x_m,y_m,t_m)$ randomly picked out from simulation data.\\

Figure \ref{fig:ModalPINN_vs_PINN}a depicts how the validation loss at the end of the training varies with the ModalPINN and the classic PINN using different numbers of degrees of freedom and different numbers of modes. Precision of the classic PINN increases with the size of the neural network. On the contrary, the ModalPINN's precision appears insensitive to the number of degrees of freedom if this number is sufficient to allow a correct representation of each mode shape. Loss convergence seems rather linked with the number of modes. Besides, for an equivalent neural network size and training time, there is an observed increase in precision up to 2 orders of magnitude for the ModalPINN. Or alternatively, to approximate the vortex shedding with the same precision as the ModalPINN with $N=3$, one would need a significantly larger classic PINN than the tested range and with a consequent increase in training time.\\

To characterise the link between precision and the number of modes in a ModalPINN, the normalised mean squared error (NMSE) is plotted. NMSE is defined as
\begin{equation}
    NMSE_q = \frac{\sum_{V_m} \left[ q(x_m,y_m,t_m) - q_m \right]^2}{\sum_{V_m} q_m^2},
\end{equation}
\noindent where $q \in \{u,v,p\}$ is the output from the ModalPINN and $q_m$ is data sampled at space-time coordinates $(x_m,y_m,t_m)$. In Figure \ref{fig:ModalPINN_vs_PINN}b, the NMSE is computed with the result of one training using successively the $k$\textsuperscript{th} first modes in addition to the steady state $\hat{q}_0$ ($N=0$). The result indicates that the precision increases with the number of modes following almost a power-law behaviour. This convergence can be compared to other previous studies like Rosenfeld et al. \cite{rosenfeld_utilization_1995} who plotted amplitude decrease of Fourier modes on a similar problem with a two order of magnitude over 6 modes. This is comparable to our result with approximately 2 orders of magnitude between 3 modes, taking into account the square norm.\\

\begin{figure}
    \centering
    \begin{tabular}{ll}
        a. & b.  \\
        \resizebox{0.47\linewidth}{!}{
\begin{tikzpicture}

\definecolor{color0}{rgb}{0.12156862745098,0.466666666666667,0.705882352941177}
\definecolor{color1}{rgb}{1,0.498039215686275,0.0549019607843137}
\definecolor{color2}{rgb}{0.172549019607843,0.627450980392157,0.172549019607843}
\definecolor{color3}{rgb}{0.83921568627451,0.152941176470588,0.156862745098039}

\begin{axis}[
log basis x={10},
log basis y={10},
tick align=outside,
tick pos=left,
x grid style={white!69.0196078431373!black},
xlabel={Number of degrees of freedom},
xmin=964.883987467383, xmax=40090.3626782465,
xmode=log,
xtick style={color=black},
xtick={10,100,1000,10000,100000,1000000},
xticklabels={\(\displaystyle 10^{1}\),\(\displaystyle 10^{2}\),\(\displaystyle 10^{3}\),\(\displaystyle 10^{4}\),\(\displaystyle 10^{5}\),\(\displaystyle 10^{6}\)},
y grid style={white!69.0196078431373!black},
ylabel={Validation Loss},
ymin=8.93558271236493e-05, ymax=0.0396101500542126,
ymode=log,
ytick style={color=black},
ytick={1e-06,1e-05,0.0001,0.001,0.01,0.1,1},
yticklabels={\(\displaystyle 10^{-6}\),\(\displaystyle 10^{-5}\),\(\displaystyle 10^{-4}\),\(\displaystyle 10^{-3}\),\(\displaystyle 10^{-2}\),\(\displaystyle 10^{-1}\),\(\displaystyle 10^{0}\)}
]
\addplot [only marks, mark=triangle*, draw=color0, fill=color0, colormap/viridis]
table{%
x                      y
2691 0.02122161
4887 0.002041131
3351 0.009729648
2103 0.028142778
1587 0.03002636
1143 0.022256328
4083 0.0033274128
9987 0.0010048815
8268 0.0011913282
5763 0.002367929
7731 0.0010242319
8823 0.0008033777
13911 0.00090244506
11223 0.0008753709
6711 0.0019999887
12531 0.00090826274
15363 0.0010327575
19308 0.0007370604
23703 0.00076555257
33843 0.00073949783
28548 0.002031639
};
\addplot [only marks, mark=square*, draw=color1, fill=color1, colormap/viridis]
table{%
x                      y
5526 0.0026439293
1566 0.0026611588
3246 0.0026139133
8406 0.0026315781
2166 0.002598781
4542 0.0026122541
};
\addplot [only marks, mark=diamond*, draw=color2, fill=color2, colormap/viridis]
table{%
x                      y
7029 0.0007287685
3339 0.0007276642
12069 0.0006969066
18459 0.00071212504
26199 0.0007024634
2241 0.00074129313
4653 0.0007176659
9891 0.0007028479
};
\addplot [only marks, mark=pentagon*, draw=color3, fill=color3, colormap/viridis]
table{%
x                      y
3852 0.00012882151
5772 0.00012492418
12252 0.00011787635
};
\end{axis}

\end{tikzpicture}} & \resizebox{0.47\linewidth}{!}{
\begin{tikzpicture}

\definecolor{color0}{rgb}{0.12156862745098,0.466666666666667,0.705882352941177}
\definecolor{color1}{rgb}{1,0.498039215686275,0.0549019607843137}
\definecolor{color2}{rgb}{0.172549019607843,0.627450980392157,0.172549019607843}
\definecolor{color3}{rgb}{0.83921568627451,0.152941176470588,0.156862745098039}

\begin{axis}[
log basis y={10},
tick align=outside,
tick pos=left,
x grid style={white!69.0196078431373!black},
xlabel={Number of modes},
xmin=-0.15, xmax=3.15,
xtick style={color=black},
xtick={0,1,2,3},
xticklabels={\(\displaystyle 0\),\(\displaystyle 1\),\(\displaystyle 2\),\(\displaystyle 3\)},
y grid style={white!69.0196078431373!black},
ylabel={Normalised Validation Error},
ymin=3.32212078151127e-05, ymax=1.63405222434409,
ymode=log,
ytick style={color=black},
ytick={1e-06,1e-05,0.0001,0.001,0.01,0.1,1,10,100},
yticklabels={\(\displaystyle 10^{-6}\),\(\displaystyle 10^{-5}\),\(\displaystyle 10^{-4}\),\(\displaystyle 10^{-3}\),\(\displaystyle 10^{-2}\),\(\displaystyle 10^{-1}\),\(\displaystyle 10^{0}\),\(\displaystyle 10^{1}\),\(\displaystyle 10^{2}\)}
]

\addplot [semithick, color0, dashed, mark=square*, mark size=2.5, mark options={solid}]
table {%
0 0.208262026309967
1 0.00963227171450853
2 0.00313658220693469
3 0.00045674224384129
};
\addplot [semithick, color1, dashed, mark=*, mark size=2.5, mark options={solid}]
table {%
0 0.0172761399298906
1 0.00110550131648779
2 0.000185004944796674
3 5.42851885256823e-05
};
\addplot [semithick, color2, dashed, mark=triangle*, mark size=2.5, mark options={solid}]
table {%
0 0.56847482919693
1 0.0247613526880741
2 0.00899676233530045
3 0.0012089591473341
};
\addplot [semithick, color3, dashed, mark=triangle*, mark size=2.5, mark options={solid,rotate=270}]
table {%
0 0.0390350595116615
1 0.00302995974197984
2 0.000227978438488208
3 0.000106982472061645
};
\end{axis}

\end{tikzpicture}}
    \end{tabular}

    \caption{With dense data: (a) Comparison of ModalPINN and PINN at given number of degrees of freedom for the same computational time (Classic PINN  \protect\tikz \protect\node[scale=0.4,regular polygon, regular polygon sides=3,fill=color0] () at (0,0){ }; , ModalPINN $N=1$ \protect\tikz \protect\node[scale=0.4,regular polygon, regular polygon sides = 4, fill=color1] () at (0,0){ }; , $2$ \protect\tikz \protect\node[scale=0.4,regular polygon, regular polygon sides = 4, fill=color2,rotate=45] () at (0,0){ }; , $3$ \protect\tikz \protect\node[scale=0.4,regular polygon, regular polygon sides = 5, fill=color3] () at (0,0){ };); (b) Evolution of normalised validation loss with the number of modes taken into account ($NMSE_{u}$ \protect\tikz \protect\node[scale=0.6,circle, fill = color1]() at (0,0){ }; , $NMSE_{v}$ \protect\tikz \protect\node[scale=0.4,regular polygon,regular polygon sides = 3, fill = color2]() at (0,0){ }; , $NMSE_{p}$ \protect\tikz \protect\node[scale=0.4,regular polygon,regular polygon sides=3, rotate=30, fill = color3]() at (0,0){ }; , average of the three \protect\tikz \protect\node[scale=0.4,regular polygon,regular polygon sides = 4, fill = color0]() at (0,0){ };)}
    \label{fig:ModalPINN_vs_PINN}
\end{figure}

\subsection{Effectiveness of modal and physical equation penalisation}

As explained in subsection \ref{subseq:LossConstruction}, two approaches can be used for training a ModalPINN with theoretical knowledge: modal and physical equations. For the incompressible flow over a cylinder, the direct method consisting in penalising mean squared residuals of equations (\ref{eq:NS_global_div_u} - \ref{eq:NS_global_f_v}) is implemented in a concise manner. The disadvantage is that a time sampling is required as well as space sampling. For a classical case, this would mean that to cover the input coordinate space with the same density in every dimension, the amount of points required would increase with the power $3/2$ compared to the 2D modal equations. \\

On the other hand, it is slightly more difficult to implement modal equations (\ref{eq:ModalEquationsDivFree} - \ref{eq:ModalEquationsY}). They usually occupy more place in memory. But in this case, only a spatial sampling is required. \\

\begin{figure}
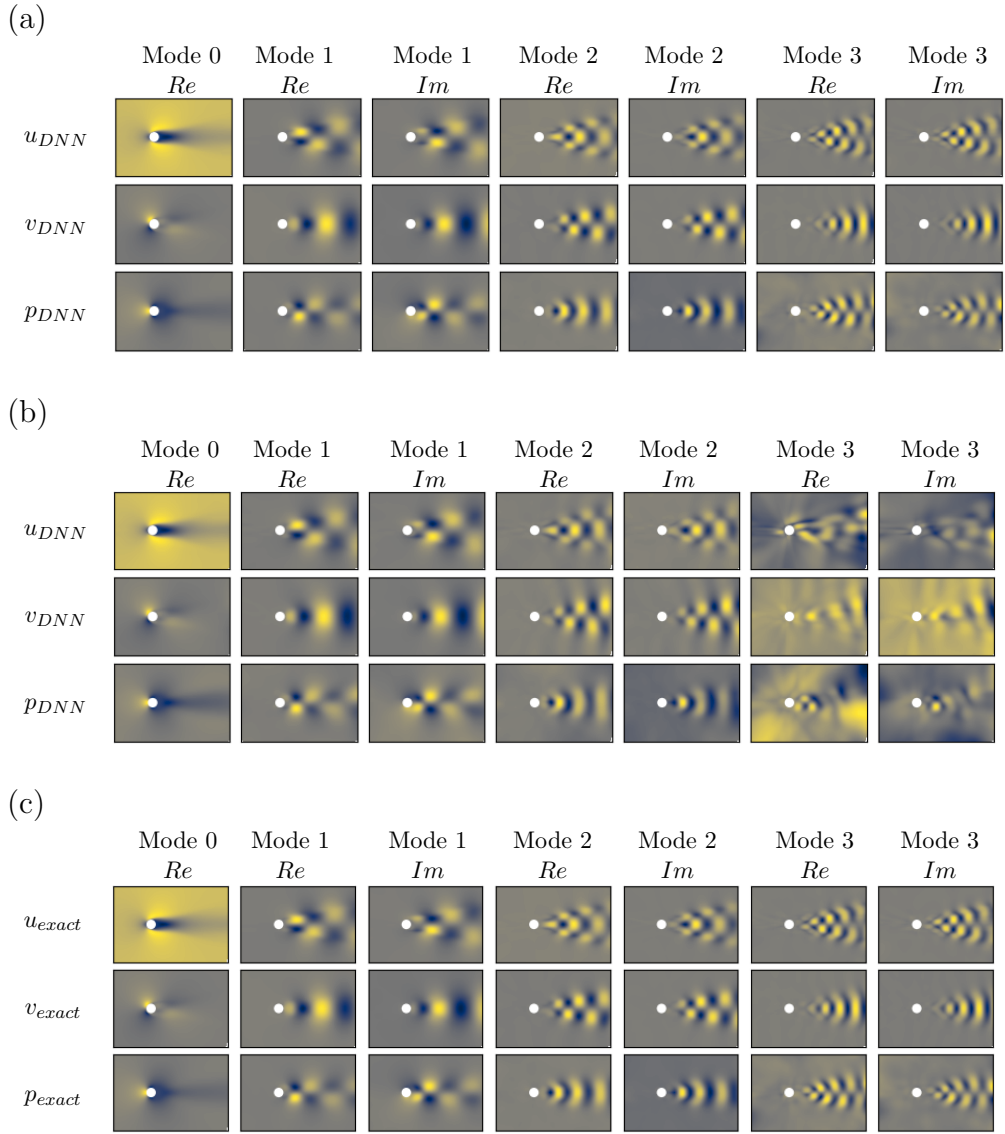


    (a) \\
    \resizebox{\linewidth}{!}{
    \input{Python_data_figure/EqsModalesVsEqsGlobalesMesDenses/figure_tikz_eqs_globales}} \\
    (b) \\
    \resizebox{\linewidth}{!}{
    \input{Python_data_figure/EqsModalesVsEqsGlobalesMesDenses/figure_tikz_eqs_modales}
    } \\
    (c) \\
    \resizebox{\linewidth}{!}{
    \input{Python_data_figure/EqsModalesVsEqsGlobalesMesDenses/figure_tikz_eqs_data}
    } \\
    \caption{Mode shapes computed with: (a) physical equations; (b) modal equations using $N_{m}=\num{5e3}$ dense data with $N=3$ modes. (c) Comparison with the mode shapes obtained directly from the complete set of reference data.}
    \label{fig:ModeShapesEqsGlobalVsEqsModal}
\end{figure}

Mode shapes obtained with both equation types are depicted in Figure \ref{fig:ModeShapesEqsGlobalVsEqsModal}. Mode shapes from physical equations in Figure \ref{fig:ModeShapesEqsGlobalVsEqsModal}a are in good agreement compared to those extracted from our reference data and plotted in Figure \ref{fig:ModeShapesEqsGlobalVsEqsModal}c. On the contrary, those obtained with modal equations in Figure \ref{fig:ModeShapesEqsGlobalVsEqsModal}b show some discrepancies starting from mode 2 with the vertical velocity $\hat{v}_2$ being poorly converged in the area downstream. Especially, the third mode did not converge for any of the three fields. Convergence of training loss is compared for both cases in Figure \ref{fig:ConvergenceHistoryModalvsPhysical}a and training with modal equations seems to reach a plateau in fewer iterations than with physical equations. Evolution of the loss function for a classical PINN is plotted in comparison in Figure \ref{fig:ConvergenceHistoryModalvsPhysical}b and depicts a slower convergence for a comparable amount of parameters. Final values of loss components $\mathcal{L}_m$ and $\mathcal{L}_{eq}$ at the end of the training are summarised in Figure \ref{fig:ConvergenceHistoryModalvsPhysical}c and it can be noted that training with physical equation resulted in a reconstruction one order of magnitude more accurate. This leads to the conclusion that computations performed on modal equations might encounter more difficulties to converge properly compared to physical equations. Also as the graph of operation is denser, optimisation is significantly slower as illustrated with the number of iterations performed with each optimiser in the same training time in Figure \ref{fig:ConvergenceHistoryModalvsPhysical}c. Thus a larger training time may be required for a similar number of iterations or targeted precision.\\

\begin{figure}
    \centering
    
    \begin{tabular}{ll}
     (a) & (b) \\
    \resizebox{0.47\linewidth}{!}{
     \input{Python_data_figure/EqsModalesVsEqsGlobalesMesDenses/Convergence_history_comparison}} & \resizebox{0.47\linewidth}{!}{
    \input{Python_data_figure/EqsModalesVsEqsGlobalesMesDenses/Convergence_history_classical_PINN}} \\
    (c) & \\
    \end{tabular} 
    \medskip
    \footnotesize
    \begin{tabular}{c|ccc}
     Type of PINN & ModalPINN & ModalPINN & Classical PINN \\
     \hline
     Training Loss & $\mathcal{L}_{eq,p} + \mathcal{L}_{m}$ & $\mathcal{L}_{eq,m} + \mathcal{L}_{m}$ & $\mathcal{L}_{eq,p} + \mathcal{L}_{m}$ \\
     Run ID & 3 & 4 & 8 \\
    $\mathcal{L}_{eq,p}$ & \num{1.4e-4} & \num{5.6e-3} & \num{9.5e-4} \\
    $\mathcal{L}_{eq,m}$ & \num{2.2} & \num{1.9e-3} & - \\
    $\mathcal{L}_{m}$ (Train.) & \num{1.1e-4} & \num{4.1e-3} & \num{6.3e-4} \\
    $\mathcal{L}_{m}$ (Valid.) & \num{1.2e-4} & \num{4.4e-3} & \num{8.1e-4} \\
     It. L-BFGS-B & \num{30233} & \num{10655} & \num{35821} \\
     It. Adam & \num{37969} & \num{1943} & \num{99980} \\
    \end{tabular}

    \caption{Comparison of the training with different loss function formulations: (a) evolution of the training error of ModalPINN with loss formulated with physical equations trained sequentially with L-BFGS-B (\protect\tikz[baseline=-0.5ex] \protect\draw[thick,dashed,color0] (0,0) -- (0.5,0);) and Adam optimiser (\protect\tikz[baseline=-0.5ex] \protect\draw[thick,dotted,blue] (0,0) -- (0.5,0);) compared with the ModalPINN formulated with modal equations using L-BFGS-B (\protect\tikz[baseline=-0.5ex] \protect\draw[thick,dash dot,color1] (0,0) -- (0.5,0);) and Adam optimiser (\protect\tikz[baseline=-0.5ex] \protect\draw[thick,solid,red] (0,0) -- (0.5,0);); (b) evolution of the training error of Classical PINN with the L-BFGS-B (\protect\tikz[baseline=-0.5ex] \protect\draw[thick,dash dot,color2] (0,0) -- (0.5,0);) and Adam optimiser (\protect\tikz[baseline=-0.5ex] \protect\draw[thick,solid,pltgreen] (0,0) -- (0.5,0);); (c) along with the final values of losses and numbers of iteration. Each run is trained with the same sequence of L-BFGS-B and Adam optimisers for an allowed training time of 10h.}
    \label{fig:ConvergenceHistoryModalvsPhysical}
\end{figure}

\subsection{Field reconstruction with data from simulated measurements}

\begin{figure}
    \centering
    \resizebox{0.6\linewidth}{!}{
    \input{Python_data_figure/Resynchro_picture/figure_tikz_resyncro}
    }
    \caption{Locations of the simulated probes with velocity data points $u$ and $v$ (\protect \tikz[baseline=-0.5ex] \protect \draw (0,0) node[cross,rotate=0] {};) and pressure sensors $p$ (\protect \tikz[baseline=-0.5ex] \protect \draw (0,0) node[cross,rotate=45,blue] {};). The practical problem of setting the time origin while sampling data at several locations is illustrated with a shift in the out-of-plane direction of a time signal (\protect \tikz[baseline=-0.5ex] \protect \draw[snake=coil,segment aspect=0,red,thick] (0,0) -- (0.6,0);).}
    \label{fig:SimulatedMeasurementSetUp}
\end{figure}

PINNs have already been proved to work well with dense information, either direct (measurement of velocity and pressure) or indirect (concentration of a passive scalar for instance) as demonstrated by Raissi et al. \cite{raissi_deep_2019}. Also they were shown to be able to infer hidden variables from equations, such as the pressure field using only velocity measurements \cite{raissi_physics-informed_2019}. This section aims at evaluating the ability of ModalPINNs to deal with very sparse and asymmetrical data distributed in a simulated experimental framework. The purpose of the following sub-sections is to quantify ModalPINN robustness when confronted with added noise and delay, which are likely to occur in an actual experiment.\\

As depicted in Figure \ref{fig:SimulatedMeasurementSetUp}, the set-up consists in 4 sections of 10 data points where a time signal of velocity $(u,v)$ is sampled ($201$ points in time covering approximately 3 periods). Such a set-up simulates an array of pitot or hot-wire measurements found in a typical laboratory experiment. The first section, which is upstream, is located at $x=-3$ starting from the centre of the cylinder. Then, the three sections downstream are respectively at $x=1$, $2$ and $3$. Then on the border of the cylinder, 30 points equally distributed on its perimeter provide information about pressure, as would do embedded pressure sensors or taps. The assumption is made that from these measurements, the fundamental frequency can be obtained with a fast Fourier transform. It is therefore fixed in the ModalPINN at the beginning of training.\\

\begin{figure}
    \centering
    \begin{tabular}{ll}
    (a) & (b) \\
    \resizebox{0.47\linewidth}{!}{\input{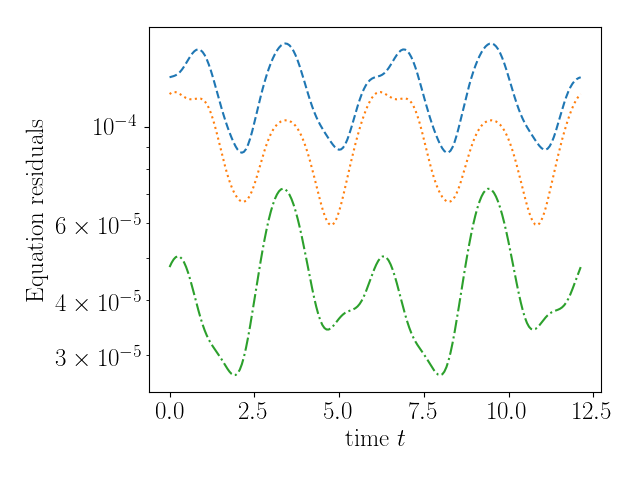}} & \resizebox{0.47\linewidth}{!}{\input{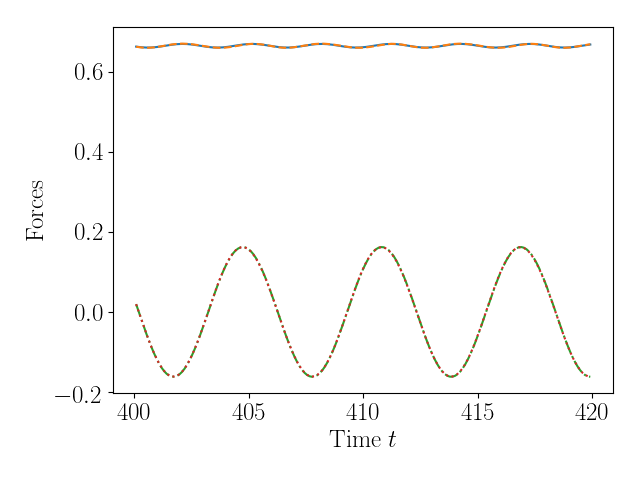}}
    \end{tabular}
    \caption{(a) Evolution in time of mean squared residuals for equations (\ref{eq:NS_global_div_u}) (\protect\tikz[baseline=-0.5ex] \protect\draw[thick,dash dot,color2] (0,0) -- (0.5,0);),  (\ref{eq:NS_global_f_u}) (\protect\tikz[baseline=-0.5ex] \protect\draw[thick,dashed,color0](0.0,0.0) -- (0.5,0.0);) and (\ref{eq:NS_global_f_v}) (\protect\tikz[baseline=-0.5ex] \protect\draw[thick,dotted,color1] (0,0) -- (0.5,0);). (b) Unsteady forces on cylinders obtained with the ModalPINN (drag $F_x$ \protect\tikz[baseline=-0.5ex] \protect\draw[thick,dashed,color1] (0,0) -- (0.5,0); and lift $F_y$ \protect\tikz[baseline=-0.5ex] \protect\draw[thick,dotted,color3] (0,0) -- (0.5,0);) and from simulation data (drag $F_x$ \protect\tikz[baseline=-0.5ex] \protect\draw[thick,solid,color0] (0,0) -- (0.5,0); and lift $F_y$ \protect\tikz[baseline=-0.5ex] \protect\draw[thick,dash dot,color2] (0,0) -- (0.5,0); ) using simulated experimental measurements. Force curves are indistinguishable.}
    \label{fig:resultsFromSimulatedMeas_Residuals_Forces}
\end{figure}

\begin{figure}
    \centering
    \includegraphics[width=0.97\linewidth]{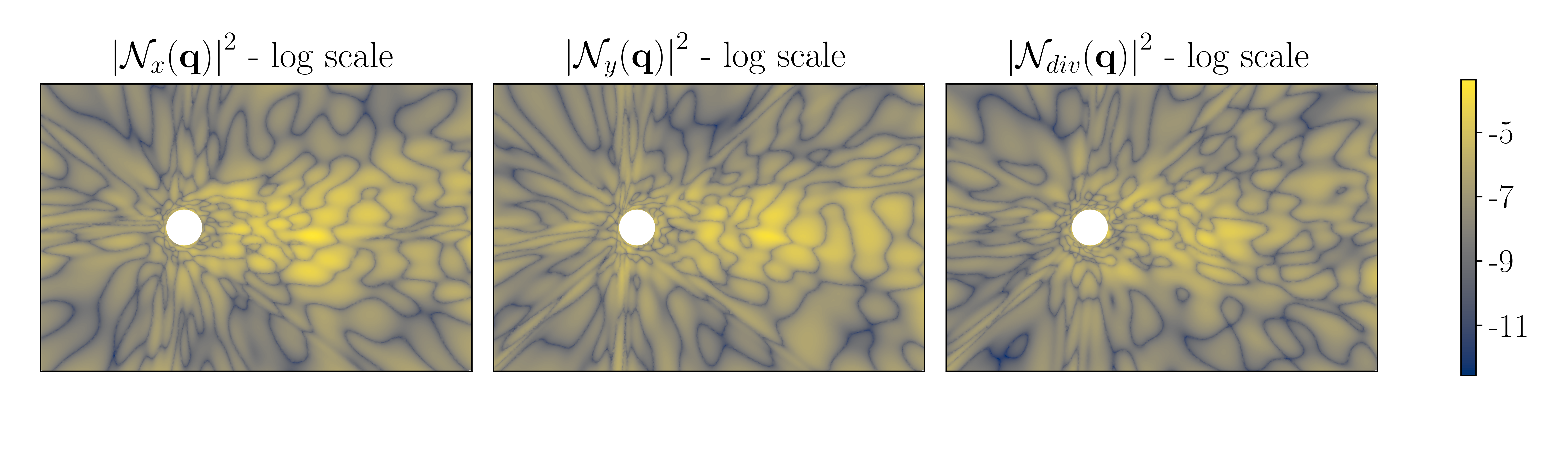}
    \caption{Space distribution of residuals on physical equations (\ref{eq:NS_global_div_u}), (\ref{eq:NS_global_f_u}) and (\ref{eq:NS_global_f_v}) at a given time-step using sparse measurements.}
    \label{fig:Distribution_residuals}
\end{figure}

Results of one training using 3 oscillating modes and physical equations are presented in Figures \ref{fig:resultsFromSimulatedMeas_NMSE_Diff}, \ref{fig:resultsFromSimulatedMeas_Residuals_Forces} and \ref{fig:Distribution_residuals}. A comparison at a given time step of predicted velocity and pressure fields with data from simulations is presented in Figure \ref{fig:resultsFromSimulatedMeas_NMSE_Diff}a and absolute difference remains small in a large area around the cylinder and in its near wake. The validation loss for different numbers of modes is shown in Figure \ref{fig:resultsFromSimulatedMeas_NMSE_Diff}b and may be compared to Figure \ref{fig:ModalPINN_vs_PINN}b, especially for the convergence of high order modes. \\

Space averaging of equations residuals is performed for several time steps in Figure \ref{fig:resultsFromSimulatedMeas_Residuals_Forces}a for each of the three equations  (\ref{eq:NS_global_div_u}-\ref{eq:NS_global_f_v}). Periodicity of these signals is a direct consequence of the enforced periodicity of ModalPINN. The residuals of both momentum equations are of the same order of magnitude, whereas the continuity equation is better satisfied. Nonetheless, these signals stay at values lower than $2\times10^{-4}$. This empirically happens to be a very acceptable value for equations residuals based on prior qualitative knowledge of cases where the exact and reconstructed flows can not be easily distinguished. The spatial distribution of residuals at a given time is presented in Figure \ref{fig:Distribution_residuals}. Error is mainly located in the wake where most of the flow unsteadiness occurs. Interestingly, the high-gradient region around the cylinder has low residuals. This is a direct consequence of the 2 zones penalisation distribution. In case of a uniform space sampling of $V_{in}$ (not shown), levels of errors are slightly higher near the cylinder border.\\

Prediction of unsteady forces are plotted alongside simulation data in Figure \ref{fig:resultsFromSimulatedMeas_Residuals_Forces}b. The horizontal and vertical forces are inferred accurately with normalised root mean square errors of \num{9.8e-4} and \num{6.1e-3} respectively. \\

\begin{figure}
    \centering
    \begin{tabular}{ll}
        (a) & (b) \\
        \resizebox{0.56\linewidth}{!}{\input{Python_data_figure/Mes_point_result/figure_tikz_comparison_in_time}} & \resizebox{0.43\linewidth}{!}{
\begin{tikzpicture}

\definecolor{color0}{rgb}{0.12156862745098,0.466666666666667,0.705882352941177}
\definecolor{color1}{rgb}{1,0.498039215686275,0.0549019607843137}
\definecolor{color2}{rgb}{0.172549019607843,0.627450980392157,0.172549019607843}
\definecolor{color3}{rgb}{0.83921568627451,0.152941176470588,0.156862745098039}

\begin{axis}[
log basis y={10},
tick align=outside,
tick pos=left,
x grid style={white!69.0196078431373!black},
xlabel={Number of modes},
xmin=-0.15, xmax=3.15,
xtick style={color=black},
xtick={0,1,2,3},
xticklabels={\(\displaystyle 0\),\(\displaystyle 1\),\(\displaystyle 2\),\(\displaystyle 3\)},
y grid style={white!69.0196078431373!black},
ylabel={Normalised Validation Error},
ymin=0.000396112842703244, ymax=1.45214044883202,
ymode=log,
ytick style={color=black},
ytick={1e-05,0.0001,0.001,0.01,0.1,1,10,100},
yticklabels={\(\displaystyle 10^{-5}\),\(\displaystyle 10^{-4}\),\(\displaystyle 10^{-3}\),\(\displaystyle 10^{-2}\),\(\displaystyle 10^{-1}\),\(\displaystyle 10^{0}\),\(\displaystyle 10^{1}\),\(\displaystyle 10^{2}\)}
]

\addplot [semithick, color0, dashed, mark=square*, mark size=3, mark options={solid}]
table {%
0 0.207379162311554
1 0.0130139421671629
2 0.00748647097498178
3 0.00731012364849448
};
\addplot [semithick, color1, dashed, mark=*, mark size=3, mark options={solid}]
table {%
0 0.0176829695701599
1 0.00143955205567181
2 0.000626487308181822
3 0.000575211481191218
};
\addplot [semithick, color2, dashed, mark=triangle*, mark size=3, mark options={solid}]
table {%
0 0.564595460891724
1 0.0321590229868889
2 0.0178603641688824
3 0.0170018896460533
};
\addplot [semithick, color3, dashed, mark=triangle*, mark size=3, mark options={solid,rotate=270}]
table {%
0 0.0398590378463268
1 0.00544324843212962
2 0.00397256063297391
3 0.00435326993465424
};
\end{axis}

\end{tikzpicture}}
    \end{tabular}
    \caption{Simulated experimental measurements:  (a) Comparison of reconstructed fields (with $N=3$) with simulated data at a given time-step $t=400\text{s}$. (b) Evolution of normalised validation loss with the number of modes ($NMSE_{u}$ \protect\tikz \protect\node[scale=0.4,regular polygon,regular polygon sides = 4, fill = color1]() at (0,0){ }; , $NMSE_{v}$ \protect\tikz \protect\node[scale=0.4,regular polygon,regular polygon sides = 3, fill = color2]() at (0,0){ }; , $NMSE_{p}$ \protect\tikz \protect\node[scale=0.4,regular polygon,regular polygon sides=3, rotate=90, fill = color3]() at (0,0){ }; , average of three \protect\tikz \protect\node[scale=0.4,regular polygon,regular polygon sides = 4, fill = color0]() at (0,0){ };)}
    \label{fig:resultsFromSimulatedMeas_NMSE_Diff}
\end{figure}

\subsection{Noise sensibility}

Measurement noise is part of the experimental process. To test the ability of the ModalPINN to deal with random perturbations, a Gaussian noise $\mathcal{N}(\mu,\sigma)$ is considered with an average $\mu=0$ and a standard deviation $\sigma$. The choice of a zero drift $\mu$ (which can be related to a kind of systematic error) and the Gaussian distribution may depend on the nature and context of the measure, as discussed by Coleman et al. \cite{coleman_experimentation_1999} for instance. The assumption is made that this is a common framework representative of real life. This noise is added to our reference data $p^{cyl},u^{probe},v^{probe}$ extracted at probe locations from numerical simulations. One obtains the new training data that are artificially flawed
\begin{equation}
    \begin{array}{rl}
        p_{noisy}^{cyl} &= p^{cyl} + \epsilon_p,  \\
        u_{noisy}^{probe} &= u^{probe} + \epsilon_u, \\
        v_{noisy}^{probe} &= v^{probe} + \epsilon_v,
    \end{array}
\end{equation}

\noindent where $\epsilon_p, \epsilon_u, \epsilon_v \sim \mathcal{N}(0,\sigma)$ at each time step and are independent from each others. Noiseless simulated probe measurements data are directly replaced by $p_{noisy}^{cyl},u_{noisy}^{probe}$ and $v_{noisy}^{probe}$ in the fitting part of the training alongside minimisation of physical equations residuals. In addition, the uncertainty on probe coordinates has been neglected in the present study but could be taken into account with a similar formalism.\\

To test the influence of noise level $\sigma$, several runs with $N=2$ oscillating modes have been carried out on a similar configuration (see section \ref{sec:TechnicalDetails} and Table 1) but with $\sigma$ taking values between \num{1e-4} and \num{1e-1}. The same noise level is added to velocity and pressure since data are physically normalised and, therefore, of order of magnitude 1. Approximately 30 jobs have been executed for each noise level so that statistical quantities that are computed can be representative. For each job, the three parts of the training loss as well as the validation loss at the end of training are presented in Figure \ref{fig:NoiseDependency}. \\

Residuals for the fitting of noisy velocity and pressure time signals are depicted in Figures \ref{fig:NoiseDependency}a and \ref{fig:NoiseDependency}b. Velocity residuals have low values (under $\sim \num{5e-4}$) for noise levels smaller than \num{1e-2} and then grow quickly with a small dispersion. Noisy pressure residuals are slightly more dispersed in the logarithmic scale for $\sigma < \num{1e-2}$ but still at low levels with an average around \num{1e-4} and a median between \num{1e-6} and \num{1e-5} before increasing with $\sigma$ as a power law. For both these noisy measurements, there is a threshold from which these fitting errors increase linearly with the square of $\sigma$. Taking into account that loss on fitting to data error is a mean square difference, this is equivalent to a linear increase of absolute measurement error with noise level. \\

Equations loss plotted in Figure \ref{fig:NoiseDependency}c shows more variability in the distribution of residuals. For nearly each level of noise there are examples of outputs that had an error of order of magnitude one, which is abnormally high and a sign of poorly converged training. This may be due to a wrong direction of optimisation or initialisation and also to the limited allocated time. Nonetheless, more than half of the results are kept at values near \num{1e-3} which, qualitatively, appears to be a very acceptable value at which usually the differences between the exact and reconstructed flow are difficult to discern for this case. Moreover, no clear increase can be noted with the level of noise for all statistical quantities of equations residuals.\\

In the end, validation loss is the quantity of interest and reveals the quality of flow reconstruction. As presented in Figure \ref{fig:NoiseDependency}d, similar conclusions as for the equations loss applied here with a small number of poorly converged results but a median that stays at low values for all noise levels. There is no discernible trend linking noise level to validation loss in the presented statistical quantities. \\

In the presence of noise, the minimisation of fitting data and the equation residuals become incompatible. Favouring the equation residuals despite the fitting error results in a large value of $\mathcal{L}_m$, and vice versa if the data are prioritised. As both terms $\mathcal{L}_m$ and $\mathcal{L}_{eq}$ have a similar weight in the total loss function $\mathcal{L}$, this choice is not encoded explicitly. From results in Figure \ref{fig:NoiseDependency}, it seems that only the minimisation of residuals drive the NN learning. Therefore the corruption of data does not affect significantly the validation error. This can be understood as a proof of robustness in the considered range of perturbation.\\

\begin{figure}
    \centering
    \begin{tabular}{ll}
        (a) & (b) \\
        \resizebox{0.47\linewidth}{!}{
\begin{tikzpicture}

\definecolor{color0}{rgb}{0.19140625,0.515625,0.53125}

\begin{axis}[
log basis x={10},
log basis y={10},
tick align=outside,
tick pos=left,
x grid style={white!69.0196078431373!black},
xlabel={Noise Level \(\displaystyle \sigma\)},
xmin=8e-05, xmax=0.12,
xmode=log,
xtick style={color=black},
xtick={1e-06,1e-05,0.0001,0.001,0.01,0.1,1,10},
xticklabels={\(\displaystyle 10^{-6}\),\(\displaystyle 10^{-5}\),\(\displaystyle 10^{-4}\),\(\displaystyle 10^{-3}\),\(\displaystyle 10^{-2}\),\(\displaystyle 10^{-1}\),\(\displaystyle 10^{0}\),\(\displaystyle 10^{1}\)},
y grid style={white!69.0196078431373!black},
ylabel={Velocity Probe Loss},
ymin=7.9297504e-05, ymax=0.0246968376,
ymode=log,
ytick style={color=black},
ytick={1e-06,1e-05,0.0001,0.001,0.01,0.1,1},
yticklabels={\(\displaystyle 10^{-6}\),\(\displaystyle 10^{-5}\),\(\displaystyle 10^{-4}\),\(\displaystyle 10^{-3}\),\(\displaystyle 10^{-2}\),\(\displaystyle 10^{-1}\),\(\displaystyle 10^{0}\)}
]
\path [draw=color0, fill=color0, opacity=0.5]
(axis cs:0.0001,9.99633605e-05)
--(axis cs:0.0001,0.000295535916)
--(axis cs:0.001,0.000290452572)
--(axis cs:0.002,0.000360183146)
--(axis cs:0.005,0.000556803325)
--(axis cs:0.01,0.000551441624)
--(axis cs:0.02,0.0011010551)
--(axis cs:0.05,0.00532439127)
--(axis cs:0.1,0.0202053306)
--(axis cs:0.1,0.0196318825)
--(axis cs:0.1,0.0196318825)
--(axis cs:0.05,0.00497846037)
--(axis cs:0.02,0.00088090961)
--(axis cs:0.01,0.000295199242)
--(axis cs:0.005,0.000147171414)
--(axis cs:0.002,0.0001082818632)
--(axis cs:0.001,0.00010242383)
--(axis cs:0.0001,9.99633605e-05)
--cycle;

\addplot [only marks, mark=+, draw=black, fill=black, colormap/viridis]
table{%
x                      y
0.001 0.0001586225
0.001 0.000103373364
0.001 0.00010333708
0.001 0.00010404588
0.01 0.0003021395
0.01 0.0005507409
0.001 0.00031150633
0.01 0.00042419892
0.01 0.0004071462
0.01 0.000419208
0.1 0.02038531
0.1 0.020040028
0.1 0.019716507
0.1 0.019853655
0.1 0.020008285
0.005 0.00018653616
0.005 0.00022920148
0.002 0.0001101228
0.002 0.00011005558
0.002 0.000108353444
0.005 0.0007169695
0.002 0.0002915948
0.002 0.00022744983
0.005 0.00014718511
0.005 0.00037929253
0.02 0.0010471551
0.02 0.001556588
0.02 0.0008846158
0.05 0.0050215134
0.02 0.0009011839
0.02 0.00088361884
0.05 0.005592542
0.0001 0.00012299938
0.05 0.0051895804
0.05 0.005064384
0.001 0.000111530426
0.0001 9.978289e-05
0.001 0.00011043694
0.0001 0.00010060248
0.05 0.0049706427
0.0001 0.0001019514
0.0001 9.912188e-05
0.0001 0.00010110331
0.0001 0.00028614045
0.002 0.00012041131
0.0001 0.0006765406
0.0001 0.0002748873
0.002 0.000112664864
0.001 0.00020931325
0.002 0.00010658141
0.001 0.000108816086
0.002 0.00033557904
0.005 0.0001504225
0.01 0.00029319664
0.01 0.00029567492
0.002 0.00010964466
0.005 0.00014907404
0.005 0.00014738298
0.005 0.00014697612
0.01 0.0003006824
0.005 0.0005246465
0.01 0.00055774814
0.01 0.00029523575
0.02 0.0013504598
0.02 0.00087462156
0.02 0.0008990457
0.02 0.0008660103
0.02 0.0008925597
0.05 0.004989964
0.05 0.00503185
0.05 0.0054492867
0.05 0.0050152643
0.05 0.005177369
0.1 0.019846587
0.1 0.020176215
0.1 0.020383302
0.1 0.02002646
0.1 0.019965576
0.01 0.0003119158
0.01 0.0003062415
0.01 0.00029914748
0.01 0.00029989367
0.01 0.00029886
0.1 0.019774118
0.1 0.02015336
0.1 0.019869316
0.1 0.019603312
0.1 0.020070296
0.1 0.019635057
0.1 0.019584754
0.1 0.019758679
0.1 0.019514924
0.1 0.020006847
0.05 0.00548893
0.05 0.005009012
0.05 0.005027362
0.05 0.0049864715
0.05 0.005077103
0.05 0.005154544
0.05 0.0050853454
0.05 0.005310514
0.05 0.005011934
0.05 0.0050106463
0.02 0.00089102704
0.02 0.00093654386
0.02 0.00089436764
0.02 0.00089661276
0.02 0.0009071887
0.02 0.0009838396
0.02 0.0010154567
0.02 0.00088826224
0.02 0.0009021841
0.02 0.0008879932
0.01 0.0005632525
0.01 0.00042219853
0.01 0.0003652737
0.005 0.00015461797
0.01 0.0002944071
0.01 0.0002998017
0.001 0.0001547299
0.005 0.000217365
0.002 0.00011577769
0.005 0.00016238193
0.002 0.00013308812
0.002 0.00011414497
0.005 0.00015502101
0.002 0.0001135266
0.005 0.00015294092
0.001 0.00012042604
0.001 0.000106966705
0.005 0.00014844682
0.0001 0.000102232996
0.005 0.0005434428
0.001 0.000102846185
0.002 0.00011008375
0.002 0.00011133427
0.001 0.00010517665
0.002 0.00011067197
0.002 0.00011109763
0.001 0.00010408412
0.005 0.00014704815
0.0001 0.000104923216
0.0001 0.00010341133
0.001 0.00010255547
0.0001 0.00010100766
0.0001 0.000103293416
0.001 0.00010189727
0.002 0.00059928244
0.005 0.00067704805
0.005 0.0007174326
0.001 0.000100032565
0.002 0.0005816201
0.0001 0.0003730532
0.0001 0.00019570759
0.001 0.00027303144
0.0001 0.00031745867
0.0001 9.9988625e-05
0.005 0.00017491264
0.005 0.00015922124
0.01 0.0002985523
0.01 0.00029487067
0.01 0.00029622292
0.005 0.00015181742
0.01 0.00029624114
0.01 0.00029913028
0.01 0.0002970785
0.01 0.0002957771
0.01 0.00029934914
0.005 0.00014510204
0.1 0.019815626
0.1 0.020023664
0.1 0.019683952
0.1 0.020012392
0.1 0.019796355
0.1 0.020185556
0.1 0.019825028
0.1 0.01978045
0.1 0.020580698
0.1 0.019815816
0.05 0.0050805924
0.05 0.004956597
0.05 0.0049802456
0.05 0.004979329
0.05 0.005003578
0.05 0.0049412865
0.05 0.0050980584
0.05 0.0050988942
0.05 0.0050254227
0.05 0.0050580287
0.02 0.0008932472
0.02 0.0010894932
0.02 0.0009015411
0.02 0.0012051122
0.02 0.0008813073
0.02 0.0009011526
0.02 0.00090086885
0.02 0.00093602185
0.02 0.0008996087
0.02 0.0008773304
0.01 0.00062827923
0.01 0.00029596323
0.005 0.00016219536
0.002 0.00015283837
0.002 0.00015959592
0.002 0.00012056399
0.005 0.00019978834
0.001 0.00012635338
0.0001 0.00014170312
0.005 0.00014965286
0.002 0.00011014654
0.002 0.00011385394
0.0001 0.0001046853
0.005 0.00015035245
0.005 0.00014982083
0.002 0.000107637636
0.001 0.00010335341
0.005 0.0004036481
0.001 0.000102851875
0.002 0.000108476765
0.001 0.0001051961
0.002 0.00067486754
0.002 0.00030836833
0.0001 0.00010086886
0.0001 9.990441e-05
0.001 0.00010071658
0.001 0.00027668802
0.001 0.0002874685
0.001 0.00028296426
0.001 0.0006255869
0.001 0.00030238886
0.002 0.00010614709
0.0001 0.00019513903
0.0001 0.00012730704
0.0001 0.00010316539
0.0001 0.00010084523
0.0001 0.000101883845
0.0001 0.00024667758
};
\addplot [only marks, mark=o, draw=red, colormap/viridis]
table{%
x                      y
0.1 0.0199297375
0.05 0.00509620970666667
0.01 0.000353614262
0.005 0.000256664781666667
0.02 0.000961500598
0.0001 0.000167370935642857
0.002 0.0001898527133
0.001 0.000169182623655172
};
\addplot [only marks, mark=square, draw=blue, colormap/viridis]
table{%
x                      y
0.1 0.0198614855
0.05 0.005029606
0.01 0.00029957542
0.005 0.000157121125
0.02 0.000900238775
0.0001 0.000103352373
0.002 0.00011369027
0.001 0.000108816086
};
\addplot [semithick, black, dashed]
table {%
0.1 0.02
0.008 0.000128
};
\addplot [red, dotted]
table {%
0.0001 0.000167370935642857
0.001 0.000169182623655172
0.002 0.0001898527133
0.005 0.000256664781666667
0.01 0.000353614262
0.02 0.000961500598
0.05 0.00509620970666667
0.1 0.0199297375
};
\addplot [blue, dotted]
table {%
0.0001 0.000103352373
0.001 0.000108816086
0.002 0.00011369027
0.005 0.000157121125
0.01 0.00029957542
0.02 0.000900238775
0.05 0.005029606
0.1 0.0198614855
};
\draw (axis cs:0.015,0.00015) node[
  scale=0.9,
  anchor=base west,
  text=black,
  rotate=0.0
]{2:1};
\end{axis}

\end{tikzpicture}} & \resizebox{0.47\linewidth}{!}{
\begin{tikzpicture}

\definecolor{color0}{rgb}{0.19140625,0.515625,0.53125}

\begin{axis}[
log basis x={10},
log basis y={10},
tick align=outside,
tick pos=left,
x grid style={white!69.0196078431373!black},
xlabel={Noise Level \(\displaystyle \sigma\)},
xmin=8e-05, xmax=0.12,
xmode=log,
xtick style={color=black},
xtick={1e-06,1e-05,0.0001,0.001,0.01,0.1,1,10},
xticklabels={\(\displaystyle 10^{-6}\),\(\displaystyle 10^{-5}\),\(\displaystyle 10^{-4}\),\(\displaystyle 10^{-3}\),\(\displaystyle 10^{-2}\),\(\displaystyle 10^{-1}\),\(\displaystyle 10^{0}\),\(\displaystyle 10^{1}\)},
y grid style={white!69.0196078431373!black},
ylabel={Pressure Loss on Cylinder},
ymin=5.00529312e-07, ymax=0.0182532912,
ymode=log,
ytick style={color=black},
ytick={1e-08,1e-07,1e-06,1e-05,0.0001,0.001,0.01,0.1,1},
yticklabels={\(\displaystyle 10^{-8}\),\(\displaystyle 10^{-7}\),\(\displaystyle 10^{-6}\),\(\displaystyle 10^{-5}\),\(\displaystyle 10^{-4}\),\(\displaystyle 10^{-3}\),\(\displaystyle 10^{-2}\),\(\displaystyle 10^{-1}\),\(\displaystyle 10^{0}\)}
]
\path [draw=color0, fill=color0, opacity=0.5]
(axis cs:0.0001,1.03151607e-06)
--(axis cs:0.0001,0.000182757355)
--(axis cs:0.001,0.000150669074)
--(axis cs:0.002,0.000274344305)
--(axis cs:0.005,0.000428092577)
--(axis cs:0.01,0.000286500968)
--(axis cs:0.02,0.000536447580000001)
--(axis cs:0.05,0.00263735956)
--(axis cs:0.1,0.0102404546)
--(axis cs:0.1,0.0096153769)
--(axis cs:0.1,0.0096153769)
--(axis cs:0.05,0.00241783436)
--(axis cs:0.02,0.000389415269)
--(axis cs:0.01,9.79063161e-05)
--(axis cs:0.005,2.55731341e-05)
--(axis cs:0.002,5.14442396e-06)
--(axis cs:0.001,2.47633768e-06)
--(axis cs:0.0001,1.03151607e-06)
--cycle;

\addplot [only marks, mark=+, draw=black, fill=black, colormap/viridis]
table{%
x                      y
0.001 1.1783618e-05
0.001 2.6516393e-06
0.001 2.6671053e-06
0.001 2.4963385e-06
0.01 9.7919845e-05
0.01 0.00027602262
0.001 0.00016379745
0.01 0.00015054167
0.01 0.00011944067
0.01 0.00011644172
0.1 0.009974154
0.1 0.009964204
0.1 0.009666484
0.1 0.00950707
0.1 0.009830659
0.005 3.8331407e-05
0.005 3.8773112e-05
0.002 5.4069487e-06
0.002 4.9061396e-06
0.002 5.8304017e-06
0.005 0.00043038017
0.002 0.00023006539
0.002 0.0001559973
0.005 2.4808565e-05
0.005 0.00033816838
0.02 0.00047816674
0.02 0.0009181308
0.02 0.00040210265
0.05 0.0024438666
0.02 0.0003955555
0.02 0.00038997844
0.05 0.0027897076
0.0001 7.002562e-06
0.05 0.002580856
0.05 0.0025043613
0.001 3.1870545e-06
0.0001 9.0519336e-07
0.001 4.029682e-06
0.0001 2.1930196e-06
0.05 0.0024186224
0.0001 5.451328e-06
0.0001 1.4441728e-06
0.0001 1.168469e-06
0.0001 0.00022403443
0.002 9.673003e-06
0.0001 0.00048097753
0.0001 0.00016506718
0.002 7.185846e-06
0.001 0.00012052144
0.002 5.2383525e-06
0.001 5.701989e-06
0.002 0.0010542355
0.005 2.5997351e-05
0.01 0.000102126134
0.01 9.949284e-05
0.002 5.1709e-06
0.005 2.714476e-05
0.005 2.5765103e-05
0.005 2.5573754e-05
0.01 9.8669254e-05
0.005 0.0004278384
0.01 0.00047806697
0.01 0.000100687685
0.02 0.0008834586
0.02 0.0004241619
0.02 0.00040184974
0.02 0.000403697
0.02 0.0004327736
0.05 0.00239371
0.05 0.0025152906
0.05 0.0030247683
0.05 0.002403556
0.05 0.0025187256
0.1 0.009946177
0.1 0.009987119
0.1 0.009819829
0.1 0.00992696
0.1 0.010071514
0.01 0.000102926984
0.01 0.00010254775
0.01 0.00010121501
0.01 0.00010137868
0.01 0.00010026984
0.1 0.009766801
0.1 0.010060758
0.1 0.009494193
0.1 0.009993763
0.1 0.010021346
0.1 0.009885001
0.1 0.010307301
0.1 0.009849242
0.1 0.009637138
0.1 0.009950108
0.05 0.0028263335
0.05 0.002410742
0.05 0.0025522127
0.05 0.0025333108
0.05 0.0024430603
0.05 0.0025254665
0.05 0.0024333263
0.05 0.002620432
0.05 0.0025345085
0.05 0.0024875563
0.02 0.00040029152
0.02 0.00044238113
0.02 0.00040031725
0.02 0.00038576213
0.02 0.00039161244
0.02 0.00047858656
0.02 0.00045305042
0.02 0.00038885573
0.02 0.00038947744
0.02 0.0004309833
0.01 0.0004218587
0.01 0.000115245195
0.01 0.00016069027
0.005 2.6576881e-05
0.01 0.000102084516
0.01 0.0001012076
0.001 1.9394076e-05
0.005 4.085686e-05
0.002 8.350249e-06
0.005 3.0749547e-05
0.002 1.2968548e-05
0.002 7.5092294e-06
0.005 2.935262e-05
0.002 6.6135744e-06
0.005 2.583456e-05
0.001 1.0859813e-05
0.001 3.6570439e-06
0.005 2.5119001e-05
0.0001 1.7904999e-06
0.005 0.00018894026
0.001 3.4151992e-06
0.002 6.0015577e-06
0.002 5.563871e-06
0.001 2.3963344e-06
0.002 5.6786057e-06
0.002 4.739396e-06
0.001 2.7712672e-06
0.005 2.5567555e-05
0.0001 1.4566228e-06
0.0001 2.3761909e-06
0.001 2.2488666e-06
0.0001 1.0604448e-06
0.0001 1.9618494e-06
0.001 2.886137e-06
0.002 0.00025932177
0.005 0.00075944705
0.005 0.00052773044
0.001 2.9043204e-06
0.002 0.00040954712
0.0001 0.000568899
0.0001 4.2265954e-05
0.001 0.00014738698
0.0001 8.806186e-05
0.0001 6.2566164e-07
0.005 3.2570537e-05
0.005 2.681512e-05
0.01 9.9609584e-05
0.01 0.00010440051
0.01 9.9271456e-05
0.005 2.7472166e-05
0.01 0.000101253165
0.01 0.00010195658
0.01 9.7737466e-05
0.01 9.6868105e-05
0.01 9.7784556e-05
0.005 2.70698e-05
0.1 0.009404538
0.1 0.010258514
0.1 0.0101611
0.1 0.010238448
0.1 0.009673487
0.1 0.009800201
0.1 0.009627411
0.1 0.009761947
0.1 0.015211076
0.1 0.009720054
0.05 0.002502394
0.05 0.0025241065
0.05 0.0024800135
0.05 0.0024985373
0.05 0.0024713364
0.05 0.002491322
0.05 0.002605829
0.05 0.0025834925
0.05 0.002468426
0.05 0.0024321005
0.02 0.00038065203
0.02 0.0004926889
0.02 0.00040535018
0.02 0.0009416394
0.02 0.0004125932
0.02 0.00039346877
0.02 0.0004017758
0.02 0.0004978908
0.02 0.00040895192
0.02 0.00040578758
0.01 0.0003808061
0.01 9.818443e-05
0.005 3.397841e-05
0.002 1.2391418e-05
0.002 1.6035292e-05
0.002 1.1885384e-05
0.005 3.5677265e-05
0.001 1.013687e-05
0.0001 1.1372738e-05
0.005 2.6150137e-05
0.002 6.683366e-06
0.002 6.702645e-06
0.0001 2.6423454e-06
0.005 2.583141e-05
0.005 2.6146388e-05
0.002 5.942382e-06
0.001 3.4738375e-06
0.005 0.00019837898
0.001 3.1855172e-06
0.002 5.2995547e-06
0.001 2.8818445e-06
0.002 0.00046510523
0.002 5.808366e-05
0.0001 1.1524195e-06
0.0001 1.18468e-06
0.001 1.985311e-06
0.001 0.00012080438
0.001 9.135956e-05
0.001 0.0002781849
0.001 0.0005114268
0.001 9.698433e-05
0.002 4.7568274e-06
0.0001 5.3562166e-05
0.0001 7.1252884e-06
0.0001 1.7852079e-06
0.0001 1.8396995e-06
0.0001 9.640157e-07
0.0001 9.211156e-05
};
\addplot [only marks, mark=o, draw=red, colormap/viridis]
table{%
x                      y
0.1 0.0100505532333333
0.05 0.00253393236666667
0.01 0.000144223530166667
0.005 0.000118101532966667
0.02 0.000467733049
0.0001 6.32315031642857e-05
0.002 9.34296487266667e-05
0.001 5.63855070517241e-05
};
\addplot [only marks, mark=square, draw=blue, colormap/viridis]
table{%
x                      y
0.1 0.0099059805
0.05 0.00250337765
0.01 0.00010166763
0.005 2.8412393e-05
0.02 0.00040556888
0.0001 2.28460525e-06
0.002 6.9442455e-06
0.001 3.6570439e-06
};
\addplot [semithick, black, dashed]
table {%
0.1 0.01
0.001 1e-06
};
\addplot [red, dotted]
table {%
0.0001 6.32315031642857e-05
0.001 5.63855070517241e-05
0.002 9.34296487266667e-05
0.005 0.000118101532966667
0.01 0.000144223530166667
0.02 0.000467733049
0.05 0.00253393236666667
0.1 0.0100505532333333
};
\addplot [blue, dotted]
table {%
0.0001 2.28460525e-06
0.001 3.6570439e-06
0.002 6.9442455e-06
0.005 2.8412393e-05
0.01 0.00010166763
0.02 0.00040556888
0.05 0.00250337765
0.1 0.0099059805
};
\draw (axis cs:0.005,2.7e-06) node[
  scale=0.9,
  anchor=base west,
  text=black,
  rotate=0.0
]{2:1};
\end{axis}

\end{tikzpicture}} \\
        (c) & (d) \\
        \resizebox{0.47\linewidth}{!}{
\begin{tikzpicture}

\definecolor{color0}{rgb}{0.19140625,0.515625,0.53125}

\begin{axis}[
log basis x={10},
log basis y={10},
tick align=outside,
tick pos=left,
x grid style={white!69.0196078431373!black},
xlabel={Noise Level \(\displaystyle \sigma\)},
xmin=8e-05, xmax=0.12,
xmode=log,
xtick style={color=black},
xtick={1e-06,1e-05,0.0001,0.001,0.01,0.1,1,10},
xticklabels={\(\displaystyle 10^{-6}\),\(\displaystyle 10^{-5}\),\(\displaystyle 10^{-4}\),\(\displaystyle 10^{-3}\),\(\displaystyle 10^{-2}\),\(\displaystyle 10^{-1}\),\(\displaystyle 10^{0}\),\(\displaystyle 10^{1}\)},
y grid style={white!69.0196078431373!black},
ylabel={Equations Loss},
ymin=0.000271508096, ymax=4.7064144,
ymode=log,
ytick style={color=black},
ytick={1e-05,0.0001,0.001,0.01,0.1,1,10,100},
yticklabels={\(\displaystyle 10^{-5}\),\(\displaystyle 10^{-4}\),\(\displaystyle 10^{-3}\),\(\displaystyle 10^{-2}\),\(\displaystyle 10^{-1}\),\(\displaystyle 10^{0}\),\(\displaystyle 10^{1}\),\(\displaystyle 10^{2}\)}
]
\path [draw=color0, fill=color0, opacity=0.5]
(axis cs:0.0001,0.000515921518)
--(axis cs:0.0001,0.0371444174)
--(axis cs:0.001,0.0100291303)
--(axis cs:0.002,0.171270636)
--(axis cs:0.005,0.0319793360000001)
--(axis cs:0.01,0.0258428708)
--(axis cs:0.02,0.0183767196)
--(axis cs:0.05,0.0147278303)
--(axis cs:0.1,0.0075833093)
--(axis cs:0.1,0.00059399913)
--(axis cs:0.1,0.00059399913)
--(axis cs:0.05,0.00041177513)
--(axis cs:0.02,0.000368687683)
--(axis cs:0.01,0.000448535573)
--(axis cs:0.005,0.000493247627)
--(axis cs:0.002,0.000529545763)
--(axis cs:0.001,0.00051552278)
--(axis cs:0.0001,0.000515921518)
--cycle;

\addplot [only marks, mark=+, draw=black, fill=black, colormap/viridis]
table{%
x                      y
0.001 0.031215528
0.001 0.00053858536
0.001 0.0017295592
0.001 0.002447654
0.01 0.0008233254
0.01 0.0049891
0.001 0.0037527608
0.01 0.001677599
0.01 0.0012391127
0.01 0.0026399544
0.1 0.008666984
0.1 0.0011785895
0.1 0.00058156275
0.1 0.0006924813
0.1 0.0014144192
0.005 0.074065275
0.005 0.0015705848
0.002 0.0004425168
0.002 0.00083410565
0.002 0.0012607285
0.005 0.009428392
0.002 0.01637891
0.002 0.0034187923
0.005 0.0053120386
0.005 0.0051103826
0.02 0.0027206326
0.02 0.019282386
0.02 0.0005423872
0.05 0.0006017681
0.02 0.00040089907
0.02 0.00057002064
0.05 0.008158272
0.0001 0.00069246243
0.05 0.0203147
0.05 0.0031991191
0.001 0.0114150075
0.0001 0.0009998897
0.001 0.0047236355
0.0001 0.0040692785
0.05 0.0005115744
0.0001 0.00043273123
0.0001 0.0014779256
0.0001 0.0005008444
0.0001 0.020259092
0.002 0.1477113
0.0001 0.0076406933
0.0001 0.07654351
0.002 0.0017970914
0.001 0.0034306038
0.002 0.00053872215
0.001 0.0015243541
0.002 3.922012
0.005 0.0012576544
0.01 0.00054964103
0.01 0.00045106313
0.002 0.0004480993
0.005 0.00039275878
0.005 0.0004999324
0.005 0.000750552
0.01 0.0013405068
0.005 0.028534766
0.01 0.005895278
0.01 0.041611202
0.02 0.01827609
0.02 0.0010276635
0.02 0.00040160445
0.02 0.00039564408
0.02 0.008803944
0.05 0.00042087576
0.05 0.82879674
0.05 0.008287304
0.05 0.0010960569
0.05 0.0020561914
0.1 0.0005909619
0.1 0.018265303
0.1 0.0051732934
0.1 0.0005943366
0.1 0.0043890225
0.01 0.06562312
0.01 0.0010088823
0.01 0.00072845834
0.01 0.0006146678
0.01 0.0012344993
0.1 0.0008000099
0.1 0.006446563
0.1 0.0005970497
0.1 0.0032507218
0.1 0.0066901776
0.1 0.00077304564
0.1 0.0006413275
0.1 0.00093229656
0.1 0.0011406016
0.1 0.0045957607
0.05 0.009231693
0.05 0.0004127479
0.05 0.0015666626
0.05 0.00041297227
0.05 0.00034816432
0.05 0.009073162
0.05 0.00090188376
0.05 0.013334344
0.05 0.0018589172
0.05 0.00042453845
0.02 0.00042144247
0.02 0.030676914
0.02 0.0009808374
0.02 0.00041795793
0.02 0.0005397116
0.02 0.0022982683
0.02 0.00390687
0.02 0.000369152
0.02 0.00035090497
0.02 0.0008983309
0.01 0.058153957
0.01 0.0009664084
0.01 0.0119315265
0.005 0.0014262622
0.01 0.0015832073
0.01 0.0004841702
0.001 0.009682661
0.005 0.0035612707
0.002 0.38330466
0.005 0.0013416592
0.002 0.09590953
0.002 0.0025056803
0.005 0.016101003
0.002 0.0034947356
0.005 0.00062606257
0.001 0.0039624674
0.001 0.0006679185
0.005 0.00043308467
0.0001 0.002070623
0.005 0.0043135546
0.001 0.0013641424
0.002 0.00053859537
0.002 0.0018592823
0.001 0.00044642045
0.002 0.0034140414
0.002 0.0017730326
0.001 0.004898156
0.005 0.0006366283
0.0001 0.000615597
0.0001 0.0020412293
0.001 0.0013149213
0.0001 0.0003707455
0.0001 0.0019480188
0.001 0.0005209792
0.002 0.009683699
0.005 0.009846367
0.005 0.011483892
0.001 0.0010000895
0.002 0.030297706
0.0001 0.0072073387
0.0001 0.004216682
0.001 0.0032900858
0.0001 0.0030788907
0.0001 0.0005812752
0.005 0.016090844
0.005 0.020899383
0.01 0.00042578756
0.01 0.00034995968
0.01 0.00042568456
0.005 0.062980466
0.01 0.00046272948
0.01 0.024090834
0.01 0.00046361054
0.01 0.00045697557
0.01 0.0008130209
0.005 0.0066935536
0.1 0.0019272303
0.1 0.007462901
0.1 0.003619836
0.1 0.0063941544
0.1 0.0023453753
0.1 0.0047262744
0.1 0.00078314915
0.1 0.0005296519
0.1 0.12439379
0.1 0.0019860072
0.05 0.0004030202
0.05 0.11569805
0.05 0.0005676965
0.05 0.00036733903
0.05 0.0011185757
0.05 0.0005793108
0.05 0.0016800358
0.05 0.014107067
0.05 0.00059726153
0.05 0.0013766449
0.02 0.002300597
0.02 0.010806952
0.02 0.00033938512
0.02 2.4528384
0.02 0.0008809729
0.02 0.0005400103
0.02 0.00055933814
0.02 0.0033238814
0.02 0.00036450883
0.02 0.0006474402
0.01 0.012820485
0.01 0.004108046
0.005 0.027368402
0.002 0.046591334
0.002 0.057755195
0.002 2.8288767
0.005 0.0021024756
0.001 0.0015847011
0.0001 1.7460356
0.005 0.00038900727
0.002 0.00073004456
0.002 0.01648152
0.0001 0.0006618278
0.005 0.0007305586
0.005 0.00071257923
0.002 0.0039871684
0.001 0.0014575481
0.005 0.23386018
0.001 0.0012279572
0.002 0.0006332017
0.001 0.00043573193
0.002 0.0088898195
0.002 0.0047644028
0.0001 0.00090138044
0.0001 0.00052238314
0.001 0.0004936971
0.001 0.0051088613
0.001 0.0052234065
0.001 0.029832646
0.001 0.008397123
0.001 0.0033286707
0.002 0.00039002293
0.0001 0.009877095
0.0001 0.0011945132
0.0001 0.14152412
0.0001 0.0022848938
0.0001 0.0010276965
0.0001 0.0030026785
};
\addplot [only marks, mark=o, draw=red, colormap/viridis]
table{%
x                      y
0.1 0.00738609592666667
0.05 0.0349167562873333
0.01 0.00826542709633333
0.005 0.0182839856706667
0.02 0.0855294382333333
0.0001 0.0729206791335714
0.002 0.253224087918667
0.001 0.00500054733586207
};
\addplot [only marks, mark=square, draw=blue, colormap/viridis]
table{%
x                      y
0.1 0.00195661875
0.05 0.0012476103
0.01 0.0011216908
0.005 0.00393741265
0.02 0.00076420655
0.0001 0.00199462405
0.002 0.00345676395
0.001 0.002447654
};
\addplot [red, dotted]
table {%
0.0001 0.0729206791335714
0.001 0.00500054733586207
0.002 0.253224087918667
0.005 0.0182839856706667
0.01 0.00826542709633333
0.02 0.0855294382333333
0.05 0.0349167562873333
0.1 0.00738609592666667
};
\addplot [blue, dotted]
table {%
0.0001 0.00199462405
0.001 0.002447654
0.002 0.00345676395
0.005 0.00393741265
0.01 0.0011216908
0.02 0.00076420655
0.05 0.0012476103
0.1 0.00195661875
};
\end{axis}

\end{tikzpicture}} & \resizebox{0.47\linewidth}{!}{        
\begin{tikzpicture}

\definecolor{color0}{rgb}{0.19140625,0.515625,0.53125}

\begin{axis}[
log basis x={10},
log basis y={10},
tick align=outside,
tick pos=left,
x grid style={white!69.0196078431373!black},
xlabel={Noise Level \(\displaystyle \sigma\)},
xmin=8e-05, xmax=0.12,
xmode=log,
xtick style={color=black},
xtick={1e-06,1e-05,0.0001,0.001,0.01,0.1,1,10},
xticklabels={\(\displaystyle 10^{-6}\),\(\displaystyle 10^{-5}\),\(\displaystyle 10^{-4}\),\(\displaystyle 10^{-3}\),\(\displaystyle 10^{-2}\),\(\displaystyle 10^{-1}\),\(\displaystyle 10^{0}\),\(\displaystyle 10^{1}\)},
y grid style={white!69.0196078431373!black},
ylabel={Validation Loss},
ymin=0.001, ymax=0.1,
ymode=log,
ytick style={color=black},
ytick={0.001,0.01,0.1},
yticklabels={\(\displaystyle 10^{-3}\),\(\displaystyle 10^{-2}\),\(\displaystyle 10^{-1}\)}
]
\path [draw=color0, fill=color0, opacity=0.5]
(axis cs:0.0001,0.00181722595)
--(axis cs:0.0001,0.0196243718)
--(axis cs:0.001,0.0170372678)
--(axis cs:0.002,0.0298128736)
--(axis cs:0.005,0.0301601368)
--(axis cs:0.01,0.0211077123)
--(axis cs:0.02,0.0176121477)
--(axis cs:0.05,0.0286771179)
--(axis cs:0.1,0.0210617573)
--(axis cs:0.1,0.00259343699)
--(axis cs:0.1,0.00259343699)
--(axis cs:0.05,0.00212976231)
--(axis cs:0.02,0.00173823928)
--(axis cs:0.01,0.0016431112)
--(axis cs:0.005,0.00167503317)
--(axis cs:0.002,0.0017165808)
--(axis cs:0.001,0.00175512918)
--(axis cs:0.0001,0.00181722595)
--cycle;

\addplot [only marks, mark=+, draw=black, fill=black, colormap/viridis]
table{%
x                      y
0.001 0.008660569
0.001 0.0018421422
0.001 0.0020232282
0.001 0.0027906275
0.01 0.001775617
0.01 0.021058764
0.001 0.020975225
0.01 0.012391648
0.01 0.010459118
0.01 0.008627019
0.1 0.028827665
0.1 0.0037311274
0.1 0.0047155293
0.1 0.0024790787
0.1 0.002833598
0.005 0.012252242
0.005 0.0056138313
0.002 0.0017393164
0.002 0.0016654041
0.002 0.0019369287
0.005 0.04489398
0.002 0.026624229
0.002 0.014563084
0.005 0.0024057
0.005 0.02306655
0.02 0.0122710075
0.02 0.038104974
0.02 0.0029205242
0.05 0.0021755458
0.02 0.0019390828
0.02 0.0017415918
0.05 0.028646594
0.0001 0.0034542999
0.05 0.01654495
0.05 0.010980628
0.001 0.0022731654
0.0001 0.001970097
0.001 0.003634121
0.0001 0.002026505
0.05 0.0022930466
0.0001 0.0018235882
0.0001 0.0017260306
0.0001 0.001625618
0.0001 0.015959777
0.002 0.005329346
0.0001 0.038331553
0.0001 0.030560441
0.002 0.0019331095
0.001 0.010793594
0.002 0.0018074231
0.001 0.003122324
0.002 0.036963213
0.005 0.0016470358
0.01 0.001573499
0.01 0.0017333366
0.002 0.0015925383
0.005 0.0016762717
0.005 0.0018448826
0.005 0.0020898955
0.01 0.0020205178
0.005 0.029365674
0.01 0.021548247
0.01 0.0015885574
0.02 0.04449328
0.02 0.002490145
0.02 0.001725217
0.02 0.0016712783
0.02 0.002634939
0.05 0.0027887186
0.05 0.0023311388
0.05 0.040286645
0.05 0.0022780183
0.05 0.014245883
0.1 0.0027692225
0.1 0.019625539
0.1 0.015559679
0.1 0.0023740546
0.1 0.020924281
0.01 0.0055547007
0.01 0.004349732
0.01 0.0018162619
0.01 0.0017186796
0.01 0.0021504574
0.1 0.002604348
0.1 0.017540526
0.1 0.003726256
0.1 0.01342753
0.1 0.00880955
0.1 0.0041593276
0.1 0.003405613
0.1 0.0028312774
0.1 0.0024952379
0.1 0.017407387
0.05 0.034898423
0.05 0.0022774462
0.05 0.005863114
0.05 0.0024421848
0.05 0.0025833522
0.05 0.017451424
0.05 0.0021320807
0.05 0.028951833
0.05 0.0071912482
0.05 0.0021477903
0.02 0.0022694485
0.02 0.007407465
0.02 0.0058862376
0.02 0.0017801215
0.02 0.0017396862
0.02 0.01008157
0.02 0.009395516
0.02 0.0018172308
0.02 0.002133251
0.02 0.0019345491
0.01 0.023050796
0.01 0.008242319
0.01 0.009791701
0.005 0.00219163
0.01 0.0024580392
0.01 0.0016302115
0.001 0.0072963014
0.005 0.007266371
0.002 0.0039271927
0.005 0.0059848316
0.002 0.005975843
0.002 0.0039073033
0.005 0.003812363
0.002 0.0025383953
0.005 0.0017343094
0.001 0.0053647626
0.001 0.0017588809
0.005 0.0016638864
0.0001 0.0022726269
0.005 0.0188903
0.001 0.002087364
0.002 0.0015878378
0.002 0.0022193433
0.001 0.0015262227
0.002 0.0021967364
0.002 0.0019396063
0.001 0.0020684048
0.005 0.0017549298
0.0001 0.0020631868
0.0001 0.0022556053
0.001 0.0018377726
0.0001 0.0019021237
0.0001 0.002201475
0.001 0.0018274329
0.002 0.040189978
0.005 0.029191319
0.005 0.044807862
0.001 0.0016443796
0.002 0.031284496
0.0001 0.028175093
0.0001 0.009555669
0.001 0.015802927
0.0001 0.015399131
0.0001 0.0021434932
0.005 0.004418658
0.005 0.0030949353
0.01 0.0017712485
0.01 0.0016445445
0.01 0.0017039601
0.005 0.0045208754
0.01 0.0021017324
0.01 0.0033886249
0.01 0.0020340753
0.01 0.0017629258
0.01 0.0023239888
0.005 0.0021648076
0.1 0.0040864837
0.1 0.022299044
0.1 0.013262574
0.1 0.012373244
0.1 0.003914869
0.1 0.01763591
0.1 0.0032320141
0.1 0.0029484485
0.1 0.04061836
0.1 0.010822508
0.05 0.0027017249
0.05 0.0029499854
0.05 0.0020034846
0.05 0.0022823967
0.05 0.0064868196
0.05 0.0022167666
0.05 0.0070975586
0.05 0.015923968
0.05 0.0020796973
0.05 0.0021088968
0.02 0.0023719901
0.02 0.015335167
0.02 0.0017558616
0.02 0.055561084
0.02 0.0017570046
0.02 0.0018456687
0.02 0.0017214926
0.02 0.008629
0.02 0.0020344695
0.02 0.0020009906
0.01 0.02797906
0.01 0.002626488
0.005 0.003802986
0.002 0.0059215045
0.002 0.008083553
0.002 0.024250718
0.005 0.008648042
0.001 0.0065589817
0.0001 0.0072868555
0.005 0.0019171634
0.002 0.0017222671
0.002 0.0032998566
0.0001 0.0022605595
0.005 0.0020175786
0.005 0.0016438566
0.002 0.0026505853
0.001 0.0021839468
0.005 0.037310302
0.001 0.0019898931
0.002 0.0018154545
0.001 0.0017401223
0.002 0.02964936
0.002 0.013611767
0.0001 0.0018284218
0.0001 0.0019368503
0.001 0.0018079255
0.001 0.016770244
0.001 0.014682916
0.001 0.018105363
0.001 0.039536983
0.001 0.016327217
0.002 0.0018015688
0.0001 0.010137325
0.0001 0.005112431
0.0001 0.0020562774
0.0001 0.002449361
0.0001 0.0018023807
0.0001 0.013162588
};
\addplot [only marks, mark=o, draw=red, colormap/viridis]
table{%
x                      y
0.1 0.0103813427566667
0.05 0.00914537876666667
0.01 0.00636252901333333
0.005 0.010389769
0.02 0.00824832813333333
0.0001 0.00755283442142857
0.002 0.0094242653
0.001 0.00748389783448276
};
\addplot [only marks, mark=square, draw=blue, colormap/viridis]
table{%
x                      y
0.1 0.00443742845
0.05 0.00274522175
0.01 0.0022372231
0.005 0.0038076745
0.02 0.00220134975
0.0001 0.0022580824
0.002 0.00297522095
0.001 0.0027906275
};
\addplot [red, dotted]
table {%
0.0001 0.00755283442142857
0.001 0.00748389783448276
0.002 0.0094242653
0.005 0.010389769
0.01 0.00636252901333333
0.02 0.00824832813333333
0.05 0.00914537876666667
0.1 0.0103813427566667
};
\addplot [blue, dotted]
table {%
0.0001 0.0022580824
0.001 0.0027906275
0.002 0.00297522095
0.005 0.0038076745
0.01 0.0022372231
0.02 0.00220134975
0.05 0.00274522175
0.1 0.00443742845
};
\end{axis}

\end{tikzpicture} }

    \end{tabular}
    \caption{Loss residuals dependency with an input noise of standard deviation $\sigma$ in measurements data. Each job result is depicted with a $+$, and for each sampling with the same level of noise, the average \protect\tikz{ \protect\draw[thick,dotted,red](-0.25,0.) -- (0.25,0.); \protect\draw[red] (0,0) circle (2pt);} is given as well as the envelope ($10-90 \%$ in \protect\tikz \protect\node[rectangle,fill={rgb,255:red,49; green,132; blue,136}] () at (0,0) { };) and the median \protect\tikz{ \protect\draw[thick,dotted,blue](-0.25,0.) -- (0.25,0.); \protect\node[scale=0.4,regular polygon,regular polygon sides = 4, draw=blue]() at (0,0){ };}. For velocity and pressure fitting error (a and b), the expected 2:1 slope for a square norm is plotted (\protect\tikz[baseline=-0.5ex] \protect\draw[thick,dashed,black](0,0) -- (0.5,0);).}
    \label{fig:NoiseDependency}
\end{figure}

\subsection{Data resynchronisation}

In the situation where a measurement is performed successively at different locations for a given recording duration with a probe, the initial condition of each measurement point is different. For a periodic phenomenon, this can be treated as an unknown phase shift for every data series, as depicted in Figure \ref{fig:SimulatedMeasurementSetUp}. To account for this phase shift in asynchronous measurements, a variable delay for each time series is determined through optimisation alongside the neural networks coefficients. To test this solution, time series of velocities from a simulated probe have been desynchronised with a random delay following a uniform law $\Delta t \sim U(0,T)$. Consequently, 40 scalar variables, one for each location, are added to the optimisation process. However, pressure measurements on the cylinder border are kept synchronised for two main reasons: the need for a constant initial phase shift to compare with validation data and because this could be carried out experimentally by parallel and synchronised pressure probes.\\

\begin{figure}
    \centering
    \begin{tabular}{cc}
            (a) & \\
    & \resizebox{0.5\linewidth}{!}{
    \input{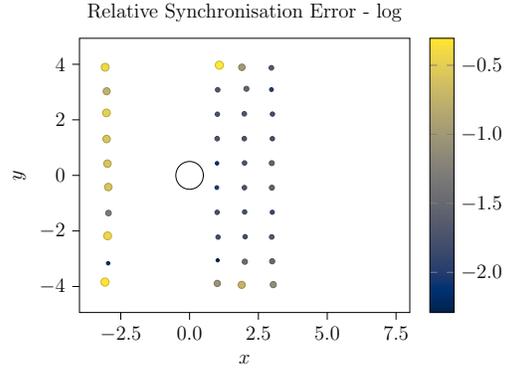}} \\
    (b) & \\
    &  \resizebox{\linewidth}{!}{
    \input{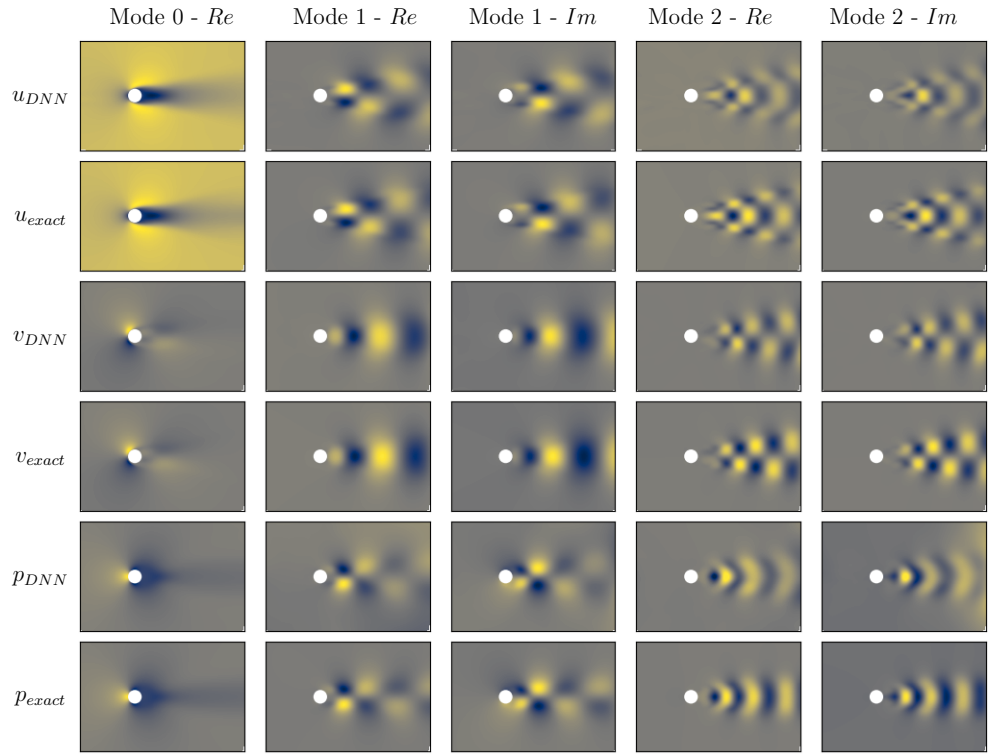}
    }
    \end{tabular}

    \caption{(a) Distribution of the residual relative error of synchronisation $ \frac{\left|\Delta t_{\text{found}} - \Delta t _{\text{exact}}\right|}{T}  < 1/2$ (depicted with a log scale). The reconstructed mode shapes are depicted at (b) and compared to the reference data from simulations.}
    \label{fig:Data_resyncro}
\end{figure}

As the array of artificially added delay for every time signal of velocity $\Delta t_{\text{exact}}$ is stored, it can be compared to the delays found through optimisation $\Delta t_{\text{found}}$ at the end of the training. The absolute difference of these two delays, centred in the interval $[-T/2,T/2]$ and then normalised by $T$ quantifies how precisely this time shift is computed. The resynchronisation error is bounded on the interval $[0,1/2]$ and its value is plotted at each probe location in Figure \ref{fig:Data_resyncro}a. The error magnitude is shown through the color contours with a logarithmic scale as well as qualitatively represented by each point size. \\

From Figure \ref{fig:Data_resyncro}a, two types of outlooks stand for re-synchronisation process. In the wake of the cylinder, the relative error on phase shift converges to approximately $1 \%$. But on the upstream sensors and at downstream probe positions that are the most distant from centre line $y=0$, residuals remain large. These points are located in areas where oscillatory phenomena are of very low amplitude. This can be seen in unsteady mode shapes. This means that there is a lack of phase information in these ranges which explains why the original phase shift can not be recovered. Fortunately, in the area of interest, phase shift is well retrieved. This can be seen in Figure \ref{fig:Data_resyncro}b where the obtained mode shapes are compared to reference data. Some differences may be noted with for instance, the second mode that has a slightly lower amplitude than the reference. But overall, there is an acceptable agreement between flow reconstruction and simulation data.\\

\section{Discussion}
\label{sec:discussion}

Differences in residual levels and convergence rates of higher frequency modes in the presence of dense data (Figure \ref{fig:ModalPINN_vs_PINN}b) or sparse data from simulated measurement (Figure \ref{fig:resultsFromSimulatedMeas_NMSE_Diff}b) can be explained with two arguments. Extrapolating flow field downstream the last measurement provided for training without any boundary condition on the outlet can be considered as an ill posed problem since there is a lack of information. The neural network simply optimises what works best with the information it has. Whereas in the dense measurement problem, there is information equally distributed in space and time, even if the space between two points can be larger than the length scale of the third mode shape. This transforms the extrapolation problem into an interpolation one, both with physical regularisation. Following these results, it would be interesting to use a PINN for flow extrapolation in the area close to a wall where measurements with hot-wire or PIV are limited \cite{kahler_uncertainty_2012,durst_experimental_2002} using the no-slip conditions in addition to the physical regularisation. This could help predict local wall shear stress with a better accuracy, which is of interest for drag estimation or for application in bio-medical applications. For instance, Arzani et al. \cite{arzani_uncovering_2021} use PINNs to estimate near-wall blood flows and wall shear stress which are linked to cardiovascular diseases.\\

The differences of convergence of higher frequency mode shapes can also be explained from a computational point of view: higher frequency modes display smaller structures than low frequency modes. In addition to the fact that these mode shapes are more complicated to approximate and thus requires larger neural networks, Navier-Stokes equations should be penalised on points that are distributed with an averaged spacing smaller than the typical wavelength. This leads to significantly increased memory requirements. In our computations this has been a limit due to the availability of RAM on the GPU (16GB in our case). For a $N=3$ modes computation, the limit in $N_{in}$ before an out of memory (OOM) error was to be found around $10^3$ points, which in that case is a small value and may not be fully adequate to thoroughly capture the steep gradients associated with small wavelength mode shapes. Besides, this can not bet fully addressed by batch processing because of the required loading time of training data after a few number of iteration. This could be overcome by using GPU with larger RAM or by splitting computation points of one optimisation iteration between different GPUs, which would require a low-level implementation.\\

In the test case of laminar vortex shedding using data at different levels of sparsity and quality, the penalisation of modal equations appeared to perform worse than physical equations penalisation, even considering that there is only a 2 dimensional input range to cover instead of 3D time-space coordinates, as illustrated with results in Figures \ref{fig:ModeShapesEqsGlobalVsEqsModal} and \ref{fig:ConvergenceHistoryModalvsPhysical}. This seems to be a consequence of non-linearities in the momentum equations that lead to sums of cross-terms at different frequencies. This makes the convergence of the solution more dependant on the number of modes and their accuracy, whereas physical equations deal with this balancing more directly and seams less affected by the truncation. Nonetheless modal equations could be of interest for linear phenomena where mode shapes might be uncoupled and computed separately, with potential applications in solid mechanics or electromagnetism for instance.\\

A variation of the ModalPINN structure could be considered, especially for linear phenomena, where $q(\mathbf{x},t) = \sum_{k=0}^N a_k \hat{q}_k(\mathbf{x})e^{ik\omega_0 t} + c.c.$ where $\hat{q}$ are mode shapes that can be computed previously and normalised $\left\| \hat{q}_k \right\|_\Omega = 1$ independently with modal equations. The modal coefficients $a_k$ can be adjusted depending on the excitation. This can ease transfer learning of once converged mode shapes to different loading configuration as in linear elasticity or electromagnetism. Also, an unknown growth exponent could be added in the presence of developing modes and optimised concurrently.\\

\section{Conclusion}
\label{sec:conclusion}

In this paper we presented an architecture of PINNs that directly approximates Fourier mode shapes. Space-time output is recovered thanks to a modal sum directly encoded in the neural network graph operation, keeping all the advantages of classical PINN while considerably reducing the required size for a similar target precision. Finally this ModalPINN structure proved to be robust to some data flaws, which makes it an efficient tool for helping academic and industrial researcher with data processing from their experimental work. This formalism can be directly extended outside of the context of fluid mechanics, as a support of digital image correlation or laser displacement measurement in solid mechanics while performing harmonic response for instance. \\

Some work remains in order to combine this technique with existing advances in PINN in order to increase robustness in the optimisation process as well as its efficiency. It goes alongside with developments in new hardware solution and implementation of libraries that better take advantage from computational resources. Finally, as we introduced this topic in the context of vortex shedding \cite{raissi_deep_2019}, extension to elastic-solid deformation of fluid-structure couplings or stability analysis could be interesting leads for future work.\\

\section{Technical details}
\label{sec:TechnicalDetails}

Training and optimisation were performed on the Graham server from Compute Canada. Each job is carried out with the same computation resources consisting in an allocation of 2 CPUs with 50 GB of RAM and a GPU Nvidia T4 (16 GB of dedicated RAM). Jobs are performed with a training limit in total duration. L-BFGS-B optimiser is used though Scipy's interface and stops when a maximum number of iteration is reached or when the difference between two iterations falls under a threshold. Only one batch of penalisation points is used during this part of the training and validation loss computation is not available. Then Adam optimisation is performed with a learning rate equal to \num{1e-5} and conducted until time limit is reached for the whole job.\\

All the scripts are written in Python using Tensorflow 1.14.1 and are available on a Github repository \cite{raynaud_modalpinn_2021}, as well as data \cite{boudina_numerical_2021} used for training. Dependencies are listed on the Github repository. Properties of every runs mentioned in the results section are summarised in Table 1.\\

\newcolumntype{P}[1]{>{\centering\arraybackslash}m{#1}}

\afterpage{
    \clearpage
    \thispagestyle{empty}
    \begin{landscape}
        \footnotesize
        \renewcommand\arraystretch{2}
        \centering 
        \resizebox{\linewidth}{!}{
        \begin{tabular}{P{0.03\linewidth}P{0.2\linewidth}P{0.08\linewidth}P{0.12\linewidth}P{0.1\linewidth}P{0.15\linewidth}P{0.1\linewidth}P{0.1\linewidth}P{0.15\linewidth}}
            \toprule
            Run ID & NN size for 1 field (activation function) & Equation type & Data type & $N_m$ & $N_{in}$ (sampling strategy) & Training Time (h) & Figures referenced & Validation Loss \\
            \midrule
            1 & $[3,W_l,W_l,W_l,1]$ $10 \leq W_l \leq 60 $ ($\sin$) & Physical & Dense & \num{5e3} & \num{5e4} (uniform) & 2 & ClassicPINN (fig. \ref{fig:ModalPINN_vs_PINN}a \protect\tikz \protect\node[scale=0.4,regular polygon, regular polygon sides=3,fill=color0] () at (0,0){ };) & From \num{7.4e-4} to \num{3.0e-2} \\
            2 & {\scriptsize $[2,W_l N^*,W_l N^*, N^*]$} $8 \leq W_l \leq 25$ and $N^*=N+1$ ($\tanh$) & Physical & Dense & \num{5e3} & \num{50e3} ($N=1$)  \num{15e3} ($N=2$)  \num{12e3} ($N=3$) (uniform) & 2 & ModalPINN (fig. \ref{fig:ModalPINN_vs_PINN}a  \protect\tikz \protect\node[scale=0.4,regular polygon, regular polygon sides = 4, fill=color1] () at (0,0){ };  \protect\tikz \protect\node[scale=0.4,regular polygon, regular polygon sides = 4, fill=color2,rotate=45] () at (0,0){ };  \protect\tikz \protect\node[scale=0.4,regular polygon, regular polygon sides = 5, fill=color3] () at (0,0){ }; and \ref{fig:ModalPINN_vs_PINN}b) & From \num{1.2e-4} to \num{1.3e-3} \\
            3 & $[2,80,80,4]$ ($\tanh$) & Physical & Dense & \num{5e3} & \num{10e3} (uniform) & 10 & Fig. \ref{fig:ModeShapesEqsGlobalVsEqsModal}a, \ref{fig:ConvergenceHistoryModalvsPhysical} & \num{1.2e-4} \\
            4 & $[2,80,80,4]$ ($\tanh$) & Modal & Dense & \num{5e3} & \num{8e3} (uniform) & 10 & Fig. \ref{fig:ModeShapesEqsGlobalVsEqsModal}b, \ref{fig:ConvergenceHistoryModalvsPhysical} & \num{4.4e-3} \\
            5 & $[2,80,80,4]$ ($\tanh$) & Physical & Simulated measurements & \num{201} time-steps per location & \num{10e3} (2 zones) & 6 & Fig. \ref{fig:resultsFromSimulatedMeas_NMSE_Diff}, \ref{fig:resultsFromSimulatedMeas_Residuals_Forces}, \ref{fig:Distribution_residuals} & \num{1.6e-3} \\
            6 & $[2,60,60,3]$ ($\tanh$) & Physical & Noisy simulated measurements & \num{201} time-steps per location & \num{15e3} (uniform) & 4 & Fig. \ref{fig:NoiseDependency} & From \num{1.5e-3} to \num{5.6e-2} \\
            7 & $[2,60,60,3]$ ($\tanh$) & Physical & Out of sync. simulated measurements & \num{201} time-steps per location & \num{20e3} (2 zones) & 10 & Fig. \ref{fig:Data_resyncro} & \num{2.8e-3} \\
            8 & $[2,60,60,60,1]$ ($\sin$) & Physical & Dense & \num{5e3} & \num{50e3} (uniform) & 10 & Fig. \ref{fig:ConvergenceHistoryModalvsPhysical}b & \num{8.1e-4} \\
            \bottomrule
        \end{tabular}}
        \label{tab:SumPropertiesRun}
        \captionof{table}{Summary of run properties which results are presented in section \ref{sec:results}. Lines 1, 2 and 6 describe a group of runs where the size of the NN (factor $W_l$) or the noise level (standard deviation $\sigma$) have been changed. $N$ denotes the number of modes chosen in ModalPINN.}
    \end{landscape}
\clearpage
}

\newpage
\section{Acknowledgements}

The authors would like to thank the participants in the Consortium on Hydraulic Machines for their support and contribution to this research project: GE Renewable Energy Canada Inc., Andritz Hydro Canada Inc., Électricité de France, Hydro-Québec, Université Laval, École Polytechnique Montréal, Vattenfall, Voith Hydro Inc. We acknowledge the support of the Natural Sciences and Engineering Council of Canada (funding reference number 507814). Our gratitude goes as well to InnovÉÉ and Calcul Canada who provided funding for this research. We also acknowledge the support of the Institut de l'Energie Trottier and of the Simulation-Based Engineering Science (SBES) program at Polytechnique Montréal.\\



\bibliographystyle{elsarticle-num}
\bibliography{references.bib}





\end{document}